\documentclass[aps,pra,twocolumn,groupedaddress,floatfix,showpacs,longbibliography]{revtex4-2}
\usepackage{graphicx}
\usepackage{amsmath}
\usepackage{color}

\begin{document}

\title{Universal Density Shift Coefficients for the\\ Thermal Conductivity and Shear Viscosity of a Unitary Fermi Gas}

\author{Xiang Li, J. Huang, and J. E. Thomas}

\affiliation{$^{1}$Department of  Physics, North Carolina State University, Raleigh, NC 27695, USA}

\date{\today}

\begin{abstract}
We measure universal temperature-independent density shifts for the thermal conductivity $\kappa_T$ and shear viscosity $\eta$, relative to the high temperature limits, for a normal phase unitary Fermi gas confined in a box potential. We show that a time-dependent kinetic theory model enables extraction of the hydrodynamic transport times $\tau_\eta$ and $\tau_\kappa$ from the time-dependent free-decay of a spatially periodic density perturbation, yielding the static transport properties and density shifts, corrected for finite relaxation times.
\end{abstract}

\maketitle

Measurements of the universal hydrodynamic transport properties of a unitary Fermi gas connect ultracold atoms to nuclear matter~\cite{NJPReview,BlochReview,UrbanReview} and provide new challenges to theoretical predictions~\cite{BruunViscousNormalDamping,RanderiaViscosity,BrabySchaeferThermalCond,DrudeBraby2011,ZwergerViscosity,LevinViscosity,DrutViscosity,
RomatschkeShear,BluhmSchaeferLocalViscosity,EnssTransport,HofmannViscosity,ZhouThermalCond}. A unitary Fermi gas is a strongly interacting, scale-invariant, quantum many-body system, created by tuning a trapped, two-component cloud near a collisional (Feshbach) resonance~\cite{OHaraScience}. At resonance, the thermodynamic and transport properties are universal functions of the density and temperature~\cite{HoUniversalThermo}, permitting parameter-free comparisons with predictions. Early measurements on expanding Fermi gas clouds with nonuniform density~\cite{CaoViscosity,JosephShearNearSF} have made way for new measurements in optical box potentials~\cite{HadzibabicBox}, where the density is nearly uniform~\cite{LorinLinearHydro,MZSound,XinHydroRelax,SecondSoundLi,HuTwoFluidPRA,MZTempWave}.
For a unitary Fermi gas, the second bulk viscosity vanishes, as predicted for scale-invariant systems~\cite{SonBulkViscosity,StringariBulk} and demonstrated in experiments on conformal symmetry~\cite{ElliottScaleInv}. Hence, in the normal phase at temperatures above the superfluid transition~\cite{KuThermo}, the hydrodynamic transport properties comprise only the shear viscosity $\eta$ and the thermal conductivity $\kappa_T$.

Remarkably, the measured shear viscosity and thermal conductivity in the normal phase appear to be fit by the simple  expressions~\cite{BluhmSchaeferLocalViscosity,XinHydroRelax,SupportOnline},
\begin{equation}
\eta =\frac{15}{32\sqrt{\pi}}\frac{(mk_BT)^{3/2}}{\hbar^2}+\alpha_{2\eta}\,\hbar n_0\,,
\label{eq:etaHighT}
\end{equation}
and
\begin{equation}
\kappa_T=\frac{15}{4}\frac{k_B}{m}\,\eta (\alpha_{2\eta}\rightarrow\alpha_{2\kappa})
\label{eq:kappaHighT}
\end{equation}
with $k_B$ the Boltzmann constant and $m$ the atom mass. The density shift coefficients $\alpha_{2\eta}$ and $\alpha_{2\kappa}$ are temperature-independent fit parameters. Here, the temperature $T$ and density $n_0$ contributions can be understood by dimensional analysis. For the shear viscosity, with a dimension of momentum/area, we expect $\eta\propto\hbar/L^3$, with $L$ a length scale. At high temperature, $L\rightarrow\lambda_T$, the thermal de Broglie wavelength $\propto T^{-1/2}$. At lower temperature, where the cloud is degenerate,  $1/L^3=n_0$. For both $\eta$ and $\kappa_T$, the leading high temperature $T^{3/2}$ dependence has been obtained by variational calculations for a unitary gas in the two-body Boltzmann limit~\cite{BruunViscousNormalDamping,BluhmSchaeferLocalViscosity,BrabySchaeferThermalCond}.  In contrast to the two-body $T^{3/2}$ coefficients, the density shift coefficients $\alpha_{2\eta}$ and $\alpha_{2\kappa}$ are unknown universal many-body parameters, which can arise in Fermi gases by Pauli blocking. In calculations of the transport times for a unitary Fermi gas, however, it has been noted that Pauli blocking appears to be nearly cancelled by in-medium scattering~\cite{EnssTransport}. Precise measurements of the density shifts therefore test the degree of this fundamental cancellation.


\begin{figure}[htb]
\includegraphics[width=3.5in]{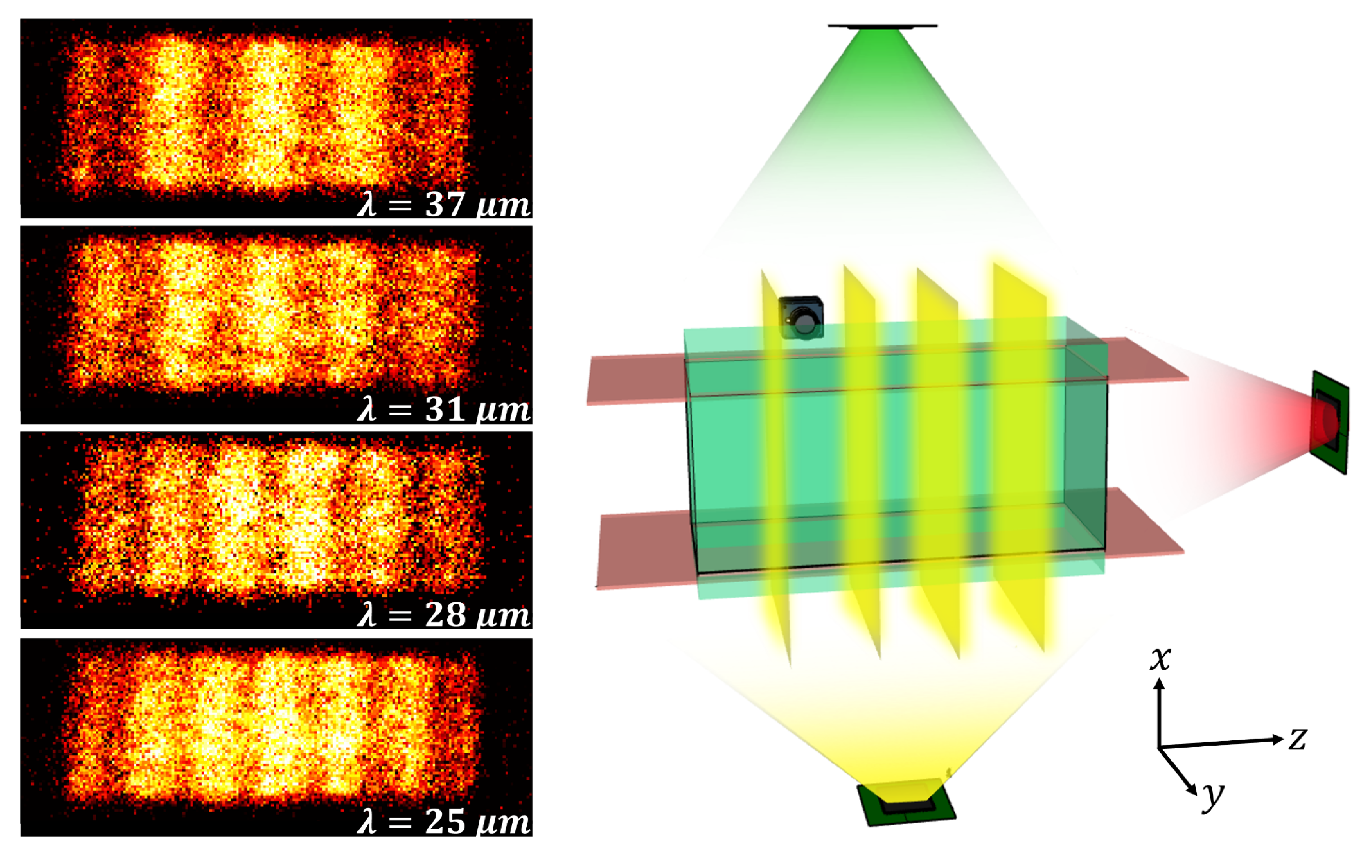}
\caption{A unitary Fermi gas is loaded into a repulsive box potential created by two digital micromirror devices DMDs (top,right).  A third DMD (bottom) generates a static spatially periodic perturbation $\delta U(z)$ with an adjustable wavelength $\lambda$, creating spatially periodic initial 1D density profiles (left). After $\delta U$ is abruptly extinguished, the dominant Fourier component exhibits an oscillatory decay (see Fig.~\ref{fig:Fourier}).
\label{fig:boxdensity}}
\end{figure}

In this work, we measure the universal temperature-independent density shifts for the static shear viscosity and thermal conductivity of a normal phase unitary Fermi gas, confined in a uniform box potential. The time-dependent free-decay of an initial spatially periodic density perturbation is observed and analyzed using a time-dependent kinetic theory model to move beyond the fast-relaxation approximation, assumed for extracting transport properties in previous experiments~\cite{MZSound,XinHydroRelax,SecondSoundLi}. The model corrects for the suppression of each transport property measured at finite frequency $\omega$, relative to the static value~\cite{Drude}.  We use this model to extract the universal hydrodynamic transport times $\tau_\eta$ and $\tau_\kappa$ from the data. The model is examined by measurements for several different perturbation wavelengths $\lambda$, which alters the frequency $\propto 1/\lambda$ and the damping rates $\propto 1/\lambda^2$. The extracted relaxation times determine the static shear viscosity and thermal conductivity, yielding the two universal density shift parameters $\alpha_{2\eta}$, $\alpha_{2\kappa}$, corrected for the finite response time over which the viscous force and heat current relax to their Navier-Stokes forms.

The experiments employ ultracold $^6$Li atoms in a balanced mixture of the two lowest hyperfine states, which are evaporatively cooled in a CO$_2$ laser trap and loaded into a box potential, Fig.~\ref{fig:boxdensity}, producing a sample of nearly uniform density $n_0$. The box comprises six sheets of blue-detuned light, created by two digital micromirror devices (DMDs)~\cite{XinHydroRelax,LorinLinearHydro} (top and right). The box potential $U_0(\mathbf{r})$ yields a rectangular density profile with typical dimensions $(x,y,z)=(52\times50\times175)\,\mu$m. The density varies  slowly in the direction of the long ($z$) axis, due to the harmonic confining potential $\propto z^2$ arising from the curvature of the bias magnetic field, which has little effect on the shorter $x$ and $y$ axes.  The typical total central density is $n_0= 3.3\times10^{11}$ atoms/cm$^3$, with the Fermi energy $\epsilon_{\!F}\equiv k_B T_F =k_B\times 0.18\,\mu$K and Fermi speed $v_F\simeq 2.25$ cm/s. The box depth $U_0\simeq 0.75\,\mu$K (see Ref.~\cite{XinHydroRelax}).

\begin{figure}[htb]
\includegraphics[width=3.0in]{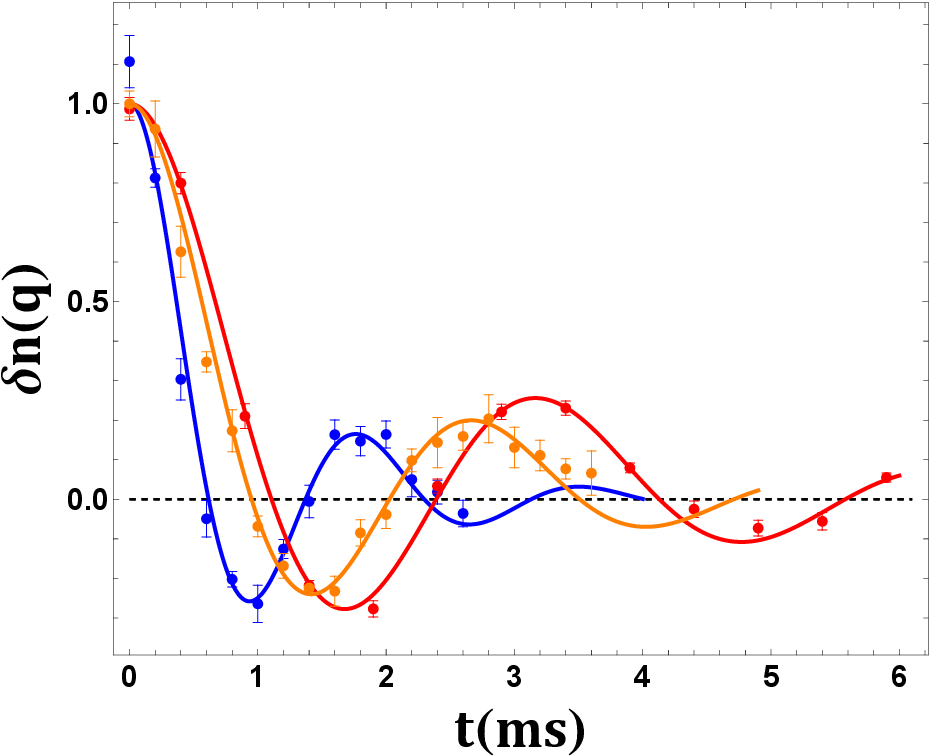}
\caption{Fourier component of the density perturbation $\delta n(q,t)$ with $q=2 \pi/\lambda$, for wavelengths $\lambda = 40.0\,\mu$m, $31.3\,\mu$m, and $22.7\,\mu$m (red, orange, blue) at reduced temperatures  $T/T_F =0.42$, $0.36$, and $0.32$, respectively. Dots: data; Curves: hydrodynamic relaxation time model.  The error bars are the standard  deviation of the mean of $\delta n(q,t)$ for 5-8 runs,  taken in random time order.
\label{fig:Fourier}}
\end{figure}

We employ a new optical system with a third DMD, Fig.~\ref{fig:boxdensity} (bottom), to independently project a static optical potential $\delta U(z)$, which is spatially periodic along one axis $z$ with an adjustable wavelength $\lambda$. This creates an initial density perturbation $\delta n(z)/n_0=A\,\sin(2\pi z/\lambda)$, where a small $A=10$\% is chosen for measurement in the hydrodynamic linear response regime~\cite{SupportOnline}. The third DMD is illuminated with a low intensity beam to utilize its full dynamic range. Once equilibrium is established, the perturbing potential is abruptly extinguished, causing an oscillatory decay of the measured density perturbation $\delta n(z,t)=n(z,t)-n_0(z)$, with $n(z,t)$ the doubly integrated 3D density. By performing a fast Fourier transform (FFT) of $\delta n(z,t)$ at each time, in a region containing an integer number (typically 3-4) of spatial periods, we obtain $\delta n(q,t)$, Fig.~\ref{fig:Fourier}.
As shown previously~\cite{XinHydroRelax}, in a fast-relaxation approximation,  $\delta n(q,t)$ contains a thermally diffusive mode ($\simeq 35$\%) that decays at a rate $\propto\kappa_T$ and an oscillating first sound mode, which decays at a rate dependent on both $\eta$ and $\kappa_T$, yielding $\eta$ and $\kappa_T$, albeit uncorrected for the finite  transport times.

To include the finite response times, we derive a relaxation model in the linear response regime by constructing four coupled equations: two describe the changes in the density $\delta n(z,t)$ and temperature $\delta T(z,t)$, and two describe the relaxation of the viscous force and heat current~\cite{BrabySchaeferThermalCond,DrudeBraby2011}. We ignore the box potential, since we measure the free-decay over time scales that avoid perturbing $\delta n(z,t)$ in the measured central region by reflections from the boundaries. After the perturbing potential is extinguished, the density change obeys~\cite{SupportOnline},
\begin{equation}
\delta\ddot{n}=c_T^2\,\partial_z^2(\delta n+\delta\tilde{T})+\delta Q_\eta\,.
\label{eq:4.5m}
\end{equation}
The $c_T^2$ term in Eq.~\ref{eq:4.5m} arises from the pressure change with $c_T$ the isothermal sound speed, and
\begin{equation}
\delta\dot{Q}_\eta+\frac{1}{\tau_\eta}\,\delta Q_\eta=\frac{4}{3}\,\frac{p}{m n_0}\,\partial_z^2\delta\dot{n}
\label{eq:4.7m}
\end{equation}
describes the relaxation of the viscous damping force~\cite{SupportOnline}. Here the pressure $p=\frac{2}{5} n\epsilon_F\!f_E(\theta)$, where the universal function $f_E(\theta)$ has been measured~\cite{KuThermo}, $n_0$ is the background density, and we have used the continuity equation to eliminate the velocity field. For fast relaxation, Eq.~\ref{eq:4.7m} yields the usual Navier-Stokes form for $\delta Q_\eta$ in Eq.~\ref{eq:4.5m} with $\eta =\tau_\eta p$ the static shear viscosity, independent of the single particle phase space distribution~\cite{SupportOnline}.

\begin{figure*}[htb]
\includegraphics[width=7.0in]{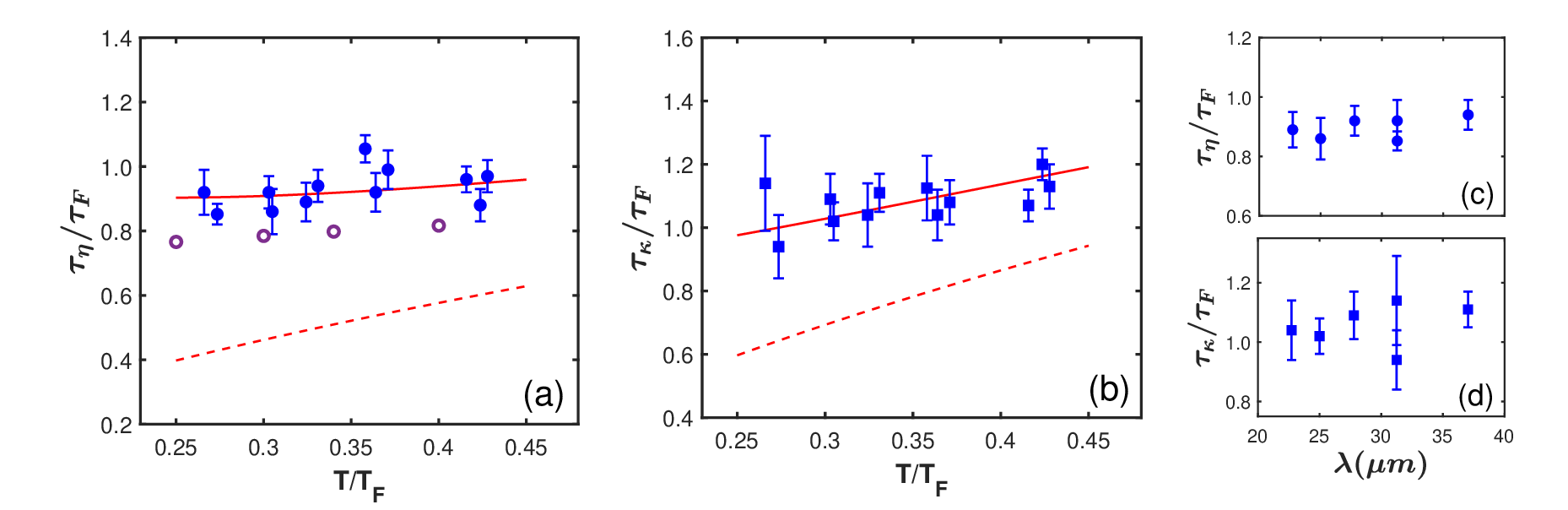}
\caption{Hydrodynamic relaxation times in units of the Fermi time $\tau_F\equiv\lambda_F/v_F=\pi\hbar/\epsilon_F$ versus reduced temperature $T/T_F$.\\ a) $\tau_\eta$ for the shear viscosity (solid blue circles). Open circles are predictions of Ref.~\cite{ZwergerViscosity}, Fig.~6 divided by $\pi$~\cite{PiFactor}. b) $\tau_\kappa$ for the thermal conductivity (blue squares). Red curves show fits with the density shift coefficients $\alpha_{2\eta}=0.45$ and $\alpha_{2\kappa}=0.22$, which are the same as obtained from the fits to Fig.~\ref{fig:transport}\,a,b.  Red dashed curves: High temperature limits, where $\alpha_{2\eta}=0$ and $\alpha_{2\kappa}=0$ and $\tau_\kappa/\tau_\eta=3/2$. (c,d) Wavelength dependence for $T/T_F\simeq 0.30$. Error bars are statistical~\cite{TransportError}.
\label{fig:taurelax}}
\end{figure*}

In Eq.~\ref{eq:4.5m}, we have defined a scaled temperature, $\delta\tilde{T}=n_0\beta\,\delta T$ with a dimension of density, where $\beta=-1/n(\partial n/\partial T)_P$ is the thermal expansivity~\cite{SupportOnline}. We find
\begin{equation}
\delta\dot{\tilde{T}}=\epsilon_{LP}\,\delta\dot{n}+\delta Q_\kappa\,.
\label{eq:8.1m}
\end{equation}
The Landau-Placzek parameter $\epsilon_{LP}\equiv c_{P_1}/c_{V_1}-1$ determines the adiabatic change in the temperature arising from the change in density. The heat capacities per particle at constant volume $c_{V_1}$ and at constant pressure $c_{P_1}$ are determined by the measured equation of state $f_E(\theta)$~\cite{KuThermo,SupportOnline} and the heat current obeys
\begin{equation}
\delta\dot{Q}_\kappa+\frac{1}{\tau_\kappa}\,\delta Q_\kappa=\frac{5}{2}\frac{k_B}{m}\frac{p}{n_0 c_{V_1}}\,\partial_z^2\delta\tilde{T}\,.
\label{eq:8.2m}
\end{equation}
For fast relaxation, Eq.~\ref{eq:8.2m} yields the usual heating rate $\delta Q_\kappa$ in Eq.~\ref{eq:8.1m} with $\kappa_T=\frac{5}{2}\frac{k_B}{m}\tau_\kappa p$ the static thermal conductivity. Here, the factor $5/2$ is dependent on a Maxwell-Boltzmann approximation for the single particle phase space distribution~\cite{SupportOnline}.

A spatial Fourier transform of Eqs.~\ref{eq:4.5m}-\ref{eq:8.2m} yields coupled linear equations for $\delta\ddot{n}(q,t)$, $\delta\dot{\tilde{T}}(q,t)$, $\delta \dot{Q}_\eta(q,t)$, and $\delta \dot{Q}_\kappa(q,t)$ with $q=2\pi/\lambda$. As the system is initially in mechanical equilibrium and isothermal, only $\delta n(q,0)\equiv A\neq 0$. Formally, the exact solutions contain four modes~\cite{SupportOnline}. However, the contributions of the two fast modes to $\delta n(q,t)$ are small ($\simeq 1$ \%)~\cite{SupportOnline} and decay quickly, since $\tau_\eta$ and $\tau_\kappa$  $\simeq 100\,\mu$s in our experiments, so they are not directly measured. The remaining thermally diffusive mode and first sound mode then dominate. The free decay of $\delta n(q,t)$ is fit by the model using the amplitude $A$, the frequency $c_Tq$, and the transport relaxation times $\tau_\eta,\tau_\kappa$ as fit parameters, instead of $A$, $c_Tq$, $\eta$, and $\kappa_T$~\cite{XinHydroRelax}. The wavelength of the perturbation and the fit frequency $c_Tq$ self-consistently determine the sound speed $c_T$ and the corresponding reduced temperature $T/T_F=\theta(c_T/v_F)$ from $f_E(\theta)$, with the Fermi speed $v_F$ given for the average central density $n_0$~\cite{SupportOnline}. Fig.~\ref{fig:Fourier} shows fits of the relaxation model (solid curves) to typical data ($\times 1/A$).

Our fitted transport times $\tau_\eta$ for the shear viscosity and $\tau_\kappa$ for the thermal conductivity are shown as functions of $\theta=T/T_F$ in Fig.~\ref{fig:taurelax}. The transport times are given in units of the Fermi time $\tau_F\equiv \lambda_F/v_F=\pi\hbar/\epsilon_F\simeq 120\,\mu$s.  We see that the fitted $\tau_\eta$ is in reasonable agreement with the predictions of Ref.~\cite{ZwergerViscosity}, Fig.~6~\cite{PiFactor}.  The wavelength dependence of $\tau_\eta$ and $\tau_\kappa$ is shown for $\theta\simeq 0.30$, demonstrating negligible $\lambda$-dependence.



\begin{figure*}[htb]
\includegraphics[width=7.0in]{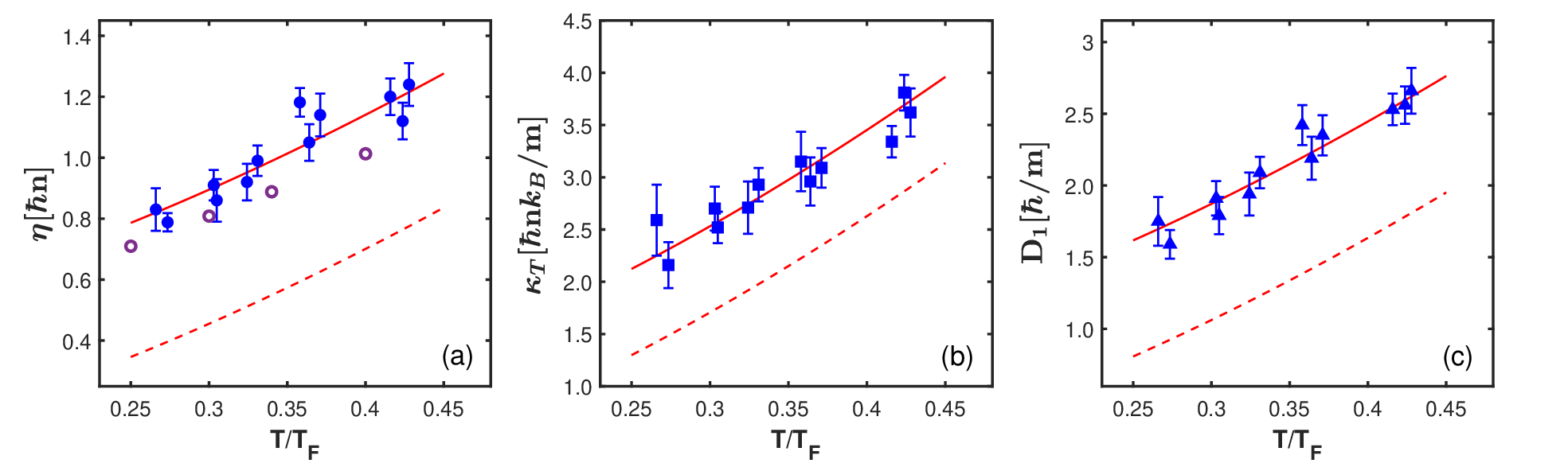}
\caption{Transport properties obtained from the measured transport times $\tau_\eta$ and $\tau_\kappa$ versus reduced temperature $\theta=T/T_F$.\\ a) Shear viscosity (Blue circles). Open circles are predictions of Ref.~\cite{ZwergerViscosity}, Fig.~7.
b) Thermal conductivity (Blue squares). c) First sound diffusivity (Blue Triangles). Red solid curves include the density shift coefficients $\alpha_{2\eta}=0.45$ in Eq.~\ref{eq:etaTheta} and $\alpha_{2\kappa}=0.22$ in Eq.~\ref{eq:kappaTheta}. Red-dashed curves are the high temperature limits, where  $\alpha_{2\eta}=0$ and $\alpha_{2\kappa}=0$. Error bars are statistical.  \label{fig:transport}}
\end{figure*}

The fitted $\tau_\eta$ determines the static shear viscosity $\eta=\tau_\eta\, p$ shown in Fig.~\ref{fig:transport}\,a (Blue circles), which is in reasonable agreement with predictions of Ref.~\cite{ZwergerViscosity}, Fig.~7 (Open circles). Eq.~\ref{eq:etaHighT}, in units of $\hbar n_0$, gives
\begin{equation}
\eta(\theta)=\alpha_{3/2}\,\theta^{3/2}+\alpha_{2\eta}\,,
\label{eq:etaTheta}
\end{equation}
where $\alpha_{3/2}=\frac{45\pi^{3/2}}{64\sqrt{2}}\simeq 2.77$~\cite{BruunViscousNormalDamping,BluhmSchaeferLocalViscosity,BrabySchaeferThermalCond}. The red curve in Fig.~\ref{fig:transport}\,a shows the fit for $\eta(\theta)$ with the density shift coefficient $\alpha_{2\eta}$ as the only fit parameter, yielding $\alpha_{2\eta}=0.45(04)$.  The red-dashed curve in Fig.~\ref{fig:transport}\,a is the high temperature limit, $\alpha_{2\eta}=0$. The red curve in Fig.~\ref{fig:taurelax}\,a shows the fit for $\tau_\eta=\eta/p$, with $p=\frac{2}{5}n\epsilon_F\!f_E(\theta)$, yielding $\alpha_{2\eta}=0.45$, the same as obtained from the fit to Fig.~\ref{fig:transport}\,a.

Similarly, the fitted $\tau_\kappa$ determines the static thermal conductivity $\kappa_T=\frac{5}{2}\frac{k_B}{m}\tau_\kappa\,p$ shown in Fig.~\ref{fig:transport}\,b (Blue squares). In the high temperature two-body Boltzmann limit, one can show that $\tau_\kappa/\tau_\eta=3/2$ for any isotropic collision cross section $d\sigma/d\Omega$, so that $\kappa_T=\frac{15}{4}\frac{k_B}{m}\,\eta$. Eq.~\ref{eq:kappaHighT}, in units of $n_0\hbar k_B/m$, gives
\begin{equation}
\kappa_T(\theta)=\frac{15}{4}\,(\alpha_{3/2}\,\theta^{3/2}+\alpha_{2\kappa})\, .
\label{eq:kappaTheta}
\end{equation}
The red curve in Fig.~\ref{fig:transport}\,b shows the fit of $\kappa_T(\theta)$ with the density shift coefficient $\alpha_{2\kappa}$ as the only fit parameter, yielding $\alpha_{2\kappa}=0.22(03)$, i.e., the shift is $15/4\times0.22$ in units of $n_0\hbar k_B/m$. The red-dashed curve in Fig.~\ref{fig:transport}\,b is the high temperature limit, $\alpha_{2\kappa}=0$. The red curve in Fig.~\ref{fig:taurelax}\,b shows the fit for $\tau_\kappa=\frac{2}{5}\frac{m}{k_B}\kappa_T/p$, yielding $\alpha_{2\kappa}=0.22$, the same as obtained from the fit to Fig.~\ref{fig:transport}\,b. We find that the measured  $\kappa_T$ are significantly smaller than predicted in~\cite{EnssTransport} and larger than predicted in~\cite{ZhouThermalCond}.

Finally, Fig.~\ref{fig:transport}\,c shows the corresponding first sound diffusivity~\cite{LandauFluids,SupportOnline} $D_1$ in units of $\hbar/m$. The fitted transport times determine $D_1[\hbar/m]=\frac{8\pi}{15}\,\frac{\tau_\eta}{\tau_F}\,f_E(\theta)+\frac{2\pi}{3}\frac{\tau_\kappa}{\tau_F}\,\theta$ (Blue triangles).  The red-solid curve gives $D_1$ in terms of the fits of Fig.~\ref{fig:transport}\,(a,b) for the static shear viscosity and thermal conductivity, $D_1=4/3\,(2.77\,\theta^{3/2}\!\!+0.45)+ (n\,k_BT/p)\,(2.77\,\theta^{3/2}\!\!+0.22)$. Here, $n\,k_BT/p=5/2\,\theta/\!f_E(\theta)$ for the unitary gas, where $p$ is the pressure~\cite{SupportOnline}. The red-dashed curve is the high temperature limit $D_1=7/3\times 2.77\,\theta^{3/2}$.

\begin{figure}[htb]
\centering
\includegraphics[width=2.75in]{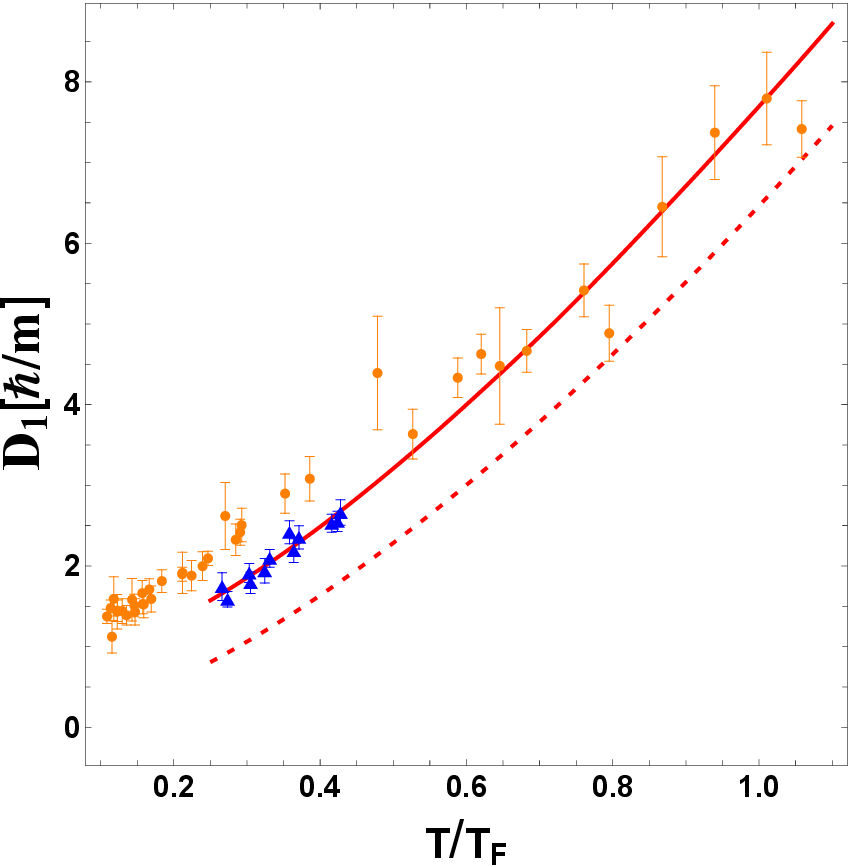}
\caption{Sound diffusivity in units of $\hbar/m$ versus reduced temperature $\theta=T/T_F$. Blue triangles: $D_1$ of Fig.~\ref{fig:transport}(c). Orange dots: Sound diffusivity data from sound attenuation~\cite{MZSound}. Red solid curve from Fig.~\ref{fig:transport}\,c: $D_1[\hbar/m]=\frac{4}{3}\,(2.77\,\theta^{3/2}\!\!+0.45)+ \frac{5}{2}\frac{\theta}{f_E(\theta)}\,(2.77\,\theta^{3/2}\!\!+0.22)$. Red-dashed curve (high temperature limit) $D_1[\hbar/m]=7/3\times 2.77\,\theta^{3/2}$. \label{fig:sounddiffusivity}}
\end{figure}

The fitted density shift coefficients for the shear viscosity and thermal conductivity are expected to be independent of reduced temperature. In Fig.~\ref{fig:sounddiffusivity}, we compare our measurements and predictions based on our measurements (red solid curve) to the first sound diffusivity measured over a wider range of reduced temperatures by sound attenuation~\cite{MZSound}. We see that the red solid curve from Fig.~\ref{fig:transport}\,c is in agreement with sound attenuation data for $T/T_F> 0.5$. However, the sound attenuation data exhibit a nearly constant upward shift relative to our hydrodynamic relaxation data where $T/T_F<0.5$, which is not yet understood.

In conclusion, we have shown that time-domain hydrodynamic relaxation experiments are well-suited for time-dependent kinetic theory models that explicitly include the transport times to determine the static transport properties. We find that the fitted transport times for the thermal current and viscous force vary slowly for reduced temperatures $0.25\leq T/T_F\leq 0.45$ and are close to one Fermi time $\lambda_F/v_F$, small, but not negligible, compared to the time scales for the oscillatory decay of the density perturbation. We obtain a ratio $\tau_\kappa/\tau_\eta\simeq 1.2$ at $T/T_F=0.4$, significantly below the high temperature Boltzmann limit, $\tau_\kappa/\tau_\eta=3/2$~\cite{EnssTransport}. The transport times determine the universal temperature-independent density shifts, providing a single parameter test of predictions for each of the static transport properties. Our measurements emphasize the need for new calculations of the leading density-dependent corrections to the hydrodynamic transport properties as well as more sophisticated time-dependent relaxation models.

We thank Thomas Sch\"{a}fer for stimulating discussions, Ilya Arakelyan for help with the optical system, Parth Patel and Martin Zwierlein for providing their sound diffusivity data, and Tilman Enss for providing predictions of the viscosity and corresponding transport time. Primary support for this research is provided by  the National Science Foundation (PHY-2006234 and PHY-2307107). Additional support is provided by the Air Force Office of Scientific Research (FA9550-22-1-0329). \\

$^*$Corresponding author: jethoma7@ncsu.edu
\newpage
%

\newpage
\widetext
\setcounter{figure}{0}
\setcounter{equation}{0}
\renewcommand{\thefigure}{S\arabic{figure}}
\renewcommand{\theequation}{S\arabic{equation}}

\appendix

\section{Supplemental Material}
In this supplemental material, we derive a time-dependent kinetic theory relaxation model for the free decay of a spatially periodic density perturbation in a normal phase unitary Fermi gas, which is confined in a box potential. Using this model, we extract the hydrodynamic transport times $\tau_\eta$ for the shear viscosity  and $\tau_\kappa$ for the thermal conductivity from free-decay data.  We show that the transport times determine the static transport properties and the corresponding universal density shift coefficients, relative to the high temperature limits, corrected for the finite response time.

\subsection{Hydrodynamic linear response for a normal fluid.}
We consider a normal phase unitary Fermi gas, which is a single component fluid with a mass density $\rho\equiv n\,m$, where $n$ is the total particle density (we assume a 50-50 mixture of two components) and $m$ is the atom mass. $\rho({\mathbf{r}},t)$ satisfies the continuity equation,
\begin{equation}
\partial_t\rho +\partial_i(\rho\,v_i)=0,
\label{eq:1.3H}
\end{equation}
where a sum over $i=x,y,z$ is implied. The mass flux (momentum density) is $\rho\,v_i$, with $v_i({\mathbf{r}},t)$ the velocity field.

Our experiments measure the response of the density in the central region of the box over short enough time scales that forces arising from the walls of the box potential can be neglected.  As the perturbing potential $\delta U=0$ during the measured evolution, the momentum density and corresponding momentum flux $\rho\,v_i v_j$ obey
\begin{equation}
\partial_t (\rho\,v_i) +\partial_j (\rho\,v_i v_j)=-\partial_i p -\partial_j p^1_{ij},
\label{eq:1.5H}
\end{equation}
where $-\partial_i p$ is the force per unit volume arising from the scalar pressure $p$ and $\partial_j p^1_{ij}$ is the viscous force per unit volume, which we determine using a kinetic theory relaxation model in S~\ref{sec:KineticTheory}.
Taking the divergence of eq.~\ref{eq:1.5H}, and using eq.~\ref{eq:1.3H}, we immediately obtain
\begin{equation}
-\partial_t^2\rho+\partial_i\partial_j (\rho\,v_i v_j)=-\partial_i^2 p - \partial_i\partial_j p^1_{ij}.
\label{eq:2.7H}
\end{equation}

In the linear response regime, the second term on the left hand side is second order in small quantities and can be dropped. Specializing to one dimension, and taking $\delta n=n-n_0$ with $n_0$ is the background density, the change in density $\delta n(z,t)$ obeys
\begin{equation}
\partial_t^2\delta n=\frac{1}{m}\partial_z^2\,\delta p\, +\frac{1}{m}\,\partial_z^2 p^1_{zz}\, .
\label{eq:3.7H}
\end{equation}
As our initial conditions are isothermal, we find the pressure change $\delta p=p-p_0$ in terms of the changes in the density $\delta n$ and  temperature $\delta T$. In this case, the pressure change can be written in the form~\cite{XinHydroRelax} $\delta p=mc_T^2(\delta n+\delta\tilde{T})$, so that
\begin{eqnarray}
\delta\ddot{n}=c_T^2\,\partial_z^2(\delta n+\delta\tilde{T})+\frac{1}{m}\,\partial_z^2 p^1_{zz}\,.
\label{eq:2.5H}
\end{eqnarray}
where $\delta\tilde{T}\equiv n_0\,\beta\,\delta{T}$ with $\beta=-1/n\,(\partial n/\partial T)_p$ the expansivity.

Next, we require the evolution equation for $\delta T$, which obeys~\cite{XinHydroRelax},
\begin{equation}
\delta\dot{T}=\epsilon_{LP}\,\frac{\delta\dot{n}}{\beta\,n_0}+\frac{\delta\dot{q}}{n_0 c_{V_1}},
\label{eq:1.9Ha}
\end{equation}
where $\epsilon_{LP}\equiv c_{P_1}/c_{V_1}-1$ is the Landau-Placzek parameter, which controls the adiabatic change in the temperature arising from the change in density.  Here,  $c_{V_1}$, $c_{P_1}$ are the heat capacities per particle. The heat flow per unit volume $\delta\dot{q}=-\partial_z J_E$, where $J_E$ is the energy current, which we determine using a kinetic theory relaxation model in S~\ref{sec:KineticTheory}.
Multiplying eq.~\ref{eq:1.9Ha} by $n_0\,\beta$, we obtain
\begin{equation}
\delta\dot{\tilde{T}}=\epsilon_{LP}\,\delta\dot{n}-\frac{\beta}{c_{V_1}}\,\partial_z J_E.
\label{eq:1.9Hb}
\end{equation}

\subsection{Kinetic theory relaxation model}
\label{sec:KineticTheory}

In this section, we derive the relaxation model equations for a normal phase unitary Fermi gas, which determine how the viscous force and heat current relax to their Navier-Stokes forms.  To proceed, we rewrite Eq.~\ref{eq:2.5H} as
\begin{eqnarray}
\delta\ddot{n}=c_T^2\,\partial_z^2(\delta n+\delta\tilde{T})+\delta Q_\eta\,,
\label{eq:2.5bS}
\end{eqnarray}
where
\begin{equation}
\delta Q_\eta\equiv\frac{1}{m}\,\partial_z^2 p^1_{zz}.
\label{eq:deltaQeta}
\end{equation}
Similarly, Eq.~\ref{eq:1.9Hb} is rewritten as
\begin{equation}
\delta\dot{\tilde{T}}=\epsilon_{LP}\,\delta\dot{n}+\delta Q_\kappa\,,
\label{eq:1.9Sd}
\end{equation}
with
\begin{equation}
\delta Q_\kappa\equiv -\frac{\beta}{c_{V_1}}\,\partial_z J_E\,.
\label{eq:deltaQkappa}
\end{equation}

We derive the evolution equations for $\delta Q_\eta$ (see Eq.~\ref{eq:4.7}) and $\delta Q_\kappa$ (see Eq.~\ref{eq:8.2}) using a  relaxation time approximation for the Boltzmann equation. In this case, the single particle phase space distribution $f(\mathbf{r}, \mathbf{v},t)$ obeys
\begin{equation}
\partial_t f+\mathbf{v}\cdot\nabla f=-\frac{1}{\tau}\,(f-f_0)\equiv -\frac{1}{\tau}\,f_1
\label{eq:1.1}
\end{equation}
with $\tau$ is the relaxation time and $f=f_0+f_1$, with $f_0$ the equilibrium distribution.

In the high-temperature Maxwell-Boltzmann limit, the equilibrium distribution is
\begin{equation}
f_0=n_0\,W_0(\mathbf{U}),
\label{eq:1.3}
\end{equation}
where $\mathbf{U}=\mathbf{v}-\mathbf{u}$ is the particle velocity relative to the stream velocity $\mathbf{u}(\mathbf{r},t)$ and $\int\! d^3\mathbf{U}\,W(\mathbf{U})=1$.
Here,
\begin{equation}
W_0(\mathbf{U})=\frac{e^{-\mathbf{U}^2/v_0^2}}{(v_0\sqrt{\pi})^3}
\label{eq:1.4}
\end{equation}
and $v_0=\sqrt{2 k_BT/m}$ is the thermal speed. In general, the background temperature $T_0$ spatially varies $T\equiv T(\mathbf{r})$. For convenience, we drop the subscript $_0$ and use $T$ for the temperature in deriving the relaxation equations.

Without specifying the phase space distribution $f$, the pressure tensor is given by
\begin{equation}
p_{ij} = m\int\!d^3\mathbf{U}\,U_i U_j\,f(\mathbf{r}, \mathbf{v},t).
\label{eq:1.5}
\end{equation}
Taking $p_{ij}=p^0_{ij}+p^1_{ij}$, the scalar pressure $p_0\equiv p$, is immediately obtained from $p^0_{ij}=\delta_{ij}\,p$ with $f=f_0$
and $i=j=x$,
\begin{equation}
p = m\int d^3\mathbf{U}\,U^2_x\,f_0.
\label{eq:1.6}
\end{equation}
Writing $\int\!d^3\mathbf{U}\,U^2_x\,f_0=n_0\langle U^2_x\rangle\equiv n_0\,\overline{U^2_x}$, we have
\begin{equation}
p = n_0 m\,\overline{U^2_x},
\label{eq:1.8}
\end{equation}
which gives $p\rightarrow n_0\,k_BT_0$ in the Maxwell-Boltsmann limit, Eq.~\ref{eq:1.4}.

\subsubsection{Shear viscosity}

We find the relaxation equation for $\delta Q_\eta$ of Eq.~\ref{eq:deltaQeta} from that of $p^1_{ij}$, assuming that the stream velocity $u(\mathbf{r},t)$ is position dependent, producing a shear stress. Here, we assume that the background temperature $T_0$ is spatially constant. To proceed, we first consider Eq.~\ref{eq:1.5} for $i\neq j$. The equilibrium distribution $f_0$ is a symmetric function of $\mathbf{U}$, so that
\begin{equation}
\int\!d^3\mathbf{U}U_i U_j\,f_0 = 0.
\label{eq:1.9}
\end{equation}
Then Eq.~\ref{eq:1.5} with $f=f_0+f_1$ and Eq.~\ref{eq:1.9} yield
\begin{equation}
p^1_{ij} = m\!\int\!d^3\mathbf{U}\,U_i U_j\,f=m\!\int\!d^3\mathbf{U}\,U_i U_j\,f_1 .
\label{eq:2.1}
\end{equation}
Multiplying Eq.~\ref{eq:1.1}  by $m\int\!d^3\mathbf{U}\,U_i U_j$ and using Eq.~\ref{eq:2.1}, we obtain
\begin{equation}
\dot{p}^1_{ij} +m\int\!d^3\mathbf{U}\,U_i U_j\,v^k\partial_k f = -\frac{1}{\tau_\eta}\,p^1_{ij},
\label{eq:2.2}
\end{equation}
where we define $\tau\equiv\tau_\eta$ for the shear viscosity. Then,
\begin{equation}
\dot{p}^1_{ij} +\frac{1}{\tau_\eta}p^1_{ij} = -I_{ij}
\label{eq:2.3}
\end{equation}
with
\begin{equation}
I_{ij} = m\int\!d^3\mathbf{U}\,U_i U_j\,v^k\partial_k f= m\int\!d^3\mathbf{U}\,U_i U_j\,v^k\frac{\partial U^l}{\partial x^k}\frac{\partial f}{\partial U^l}.
\label{eq:2.4}
\end{equation}
For fast relaxation, where $f_1\rightarrow \tau_\eta v_k\partial_kf$ in Eq.~\ref{eq:1.1},  we see that $p^1_{ij}\simeq\tau_\eta I_{ij}$ is already first order in $\tau_\eta$, so that we can take $f\rightarrow f_0$ in Eq.~\ref{eq:2.4}. Using $\mathbf{U} =\mathbf{v} -\mathbf{u}$,
\begin{equation}
\frac{\partial U^l}{\partial x^k} = -\frac{\partial u^l}{\partial x^k}.
\label{eq:2.5}
\end{equation}
Then with $v_k=U_k+u_k$ in Eq.~\ref{eq:2.4}, we obtain
\begin{equation}
I_{ij} = -m\frac{\partial u^l}{\partial x^k}\int\!d^3\mathbf{U}\,U_i U_j (U_k+u_k)\frac{\partial f_0}{\partial U^l}.
\label{eq:2.6}
\end{equation}
Integrating by parts, we then have
\begin{equation}
I_{ij}= m\frac{\partial u^l}{\partial x^k}\int\!d^3\mathbf{U}\,\frac{\partial}{\partial U^l}[U_i U_j (U_k+u_k)]\,f_0,
\label{eq:2.7}
\end{equation}
Using $\partial U_i/\partial U_l=\delta_{il}$ and defining $\int d^3\mathbf{U}\,g(\mathbf{U})f_0(\mathbf{U})=n_0\langle g(\mathbf{U})\rangle$,  Eq.~\ref{eq:2.7} can be rewritten as
\begin{equation}
I_{ij}=m\frac{\partial u^l}{\partial x^k}\,n_0\left\{\delta_{il}\langle U_j(U_k+u_k)\rangle +\delta_{jl}\langle U_i(U_k+u_k)\rangle+\delta_{kl}\langle U_i U_j\rangle\right\}
\label{eq:2.8}
\end{equation}
Eq.~\ref{eq:2.8} is simplified with
\begin{equation}
n_0 \langle U_j(U_k+u_k)\rangle=\int\!d^3\mathbf{U}\,U_j(U_k+u_k)\,f_0=n_0\,\delta_{jk}\overline{U^2_x},
\label{eq:3.1}
\end{equation}
where the $u_k$ term vanishes since $f_0$ is symmetric in $U_j$. Hence, we can write $\langle U_iU_j\rangle = \delta_{ij}\overline{U^2_x}$. With $p = n_0 m\,\overline{U^2_x}$ from Eq.~\ref{eq:1.8},
we  have
\begin{equation}
I_{ij} = p\,\frac{\partial u^l}{\partial x^k}\,\{\delta_{il}\delta_{jk}+\delta_{jl}\delta_{ik}+\delta_{kl}\delta_{ij}\}\,.
\label{eq:3.3}
\end{equation}
Carrying out the sums over repeated indices, we obtain
\begin{equation}
I_{ij}= p\left(\frac{\partial u^i}{\partial x^j}+\frac{\partial u^j}{\partial x^i}+\delta_{ij}\nabla\cdot\mathbf{u}\right),
\label{eq:3.4}
\end{equation}
where the $\delta_{ij}$ term vanishes for $i\neq j$.

To determine $p^1_{ij}$ for all $i,j$, we consider the symmetric second rank traceless pressure tensor,
\begin{equation}
p^1_{ij} \rightarrow  m\!\int\!d^3\mathbf{U}\,\left(U_i U_j-\frac{1}{3}\delta_{ij}\mathbf{U}^2\right)\,f_1\,.
\label{eq:3.5}
\end{equation}
where the $f_0$ part of $f$ vanishes because it is scalar function of $\mathbf{U}$. Since $\mathbf{U}^2=Tr\{U_iU_j\}$, evaluating Eq.~\ref{eq:3.5} just changes $I_{ij}$ in Eq.~\ref{eq:2.3} and Eq.~\ref{eq:3.4} to
\begin{equation}
I_{ij}\rightarrow I_{ij}-\frac{1}{3}\delta_{ij}Tr\{I_{ij}\}\equiv p\,\sigma_{ij}\,.
\label{eq:3.6}
\end{equation}
The $\delta_{ij}$ term in Eq.~\ref{eq:3.4} makes no contribution to the symmetric traceless tensor, yielding
\begin{equation}
\sigma_{ij} = \frac{\partial u^i}{\partial x^j}+\frac{\partial u^j}{\partial x^i} -\frac{2}{3}\,\delta_{ij}\nabla\cdot\mathbf{u}\,.
\label{eq:3.7}
\end{equation}

With Eqs.~\ref{eq:3.6}~and~\ref{eq:3.7}, Eq.~\ref{eq:2.3} determines the relaxation equation for the shear stress tensor,
\begin{equation}
\dot{p}^1_{ij} + \frac{1}{\tau_\eta}\,p^1_{ij}=-p\,\sigma_{ij}\,.
\label{eq:4.1}
\end{equation}
For small $\tau_\eta$, $\dot{p}^1_{ij}\ll p^1_{ij}/\tau_\eta$,  we see that
\begin{equation}
p^1_{ij}\rightarrow -\tau_\eta p\,\delta_{ij} = -\eta\,\sigma_{ij},
\label{eq:4.2}
\end{equation}
where the static shear viscosity is
\begin{equation}
\eta = \tau_\eta\, p.
\label{eq:4.3}
\end{equation}

Now we can evaluate the relaxation equation for $\delta Q_\eta$ of Eq.~\ref{eq:deltaQeta}.  We evaluate Eq.~\ref{eq:3.7} for $\sigma_{zz}$ and eliminate the velocity field using current conservation Eq.~\ref{eq:1.3H}. To first order in small quantities,  $\delta\dot{n}+n_0\,\partial_zv_z=0$, yielding
\begin{equation}
\sigma_{zz} = \frac{4}{3}\,\partial_z v_z=-\frac{4}{3}\,\frac{\delta\dot{n}}{n_0}\,.
\label{eq:4.6}
\end{equation}
We find $\delta\dot{Q_{\eta}}$ using Eq.~\ref{eq:deltaQeta}, $\delta Q_\eta\equiv\frac{1}{m}\,\partial_z^2 p^1_{zz}$. With Eq.~\ref{eq:4.6} and Eq.~\ref{eq:4.1} we obtain finally,
\begin{equation}
\delta\dot{Q_{\eta}}+\frac{1}{\tau_{\eta}}\delta Q_{\eta}=\frac{4}{3}\,\frac{p}{mn_0}\,\partial^2_z\delta\dot{n}\,.
\label{eq:4.7}
\end{equation}
For fast relaxation, $\delta\dot{Q_{\eta}}\ll \delta Q_{\eta}/\tau_{\eta}$, we find,
\begin{equation}
\delta Q_{\eta}\rightarrow\frac{4}{3}\,\frac{\tau_{\eta}p}{mn_0}\,\partial^2_z\delta\dot{n} = \frac{4}{3}\,\frac{\eta}{mn_0}\,\partial^2_z\delta\dot{n}\,.
\label{eq:4.8}
\end{equation}
In this limit, Eq.~\ref{eq:2.5bS} with Eq.~\ref{eq:4.8} reproduces the Navier-Stokes form used in Ref.~\cite{XinHydroRelax}. We note that Eq.~\ref{eq:4.3} for $\eta$  and Eq.~\ref{eq:4.7} for $\delta\dot{Q_{\eta}}$ are
{\it independent} of the form of the single particle phase space distribution, $f_0(\mathbf{r},\mathbf{v})$, which has not been explicitly used to obtain the background pressure $p$ from Eq.~\ref{eq:1.8}.

\subsubsection{Thermal conductivity}

Next, we find the relaxation equation for $\delta Q_\kappa$ of Eq.~\ref{eq:deltaQkappa} from that of the 1D energy current, $J_E$,
\begin{equation}
J_E = \int d^3\mathbf{v}\,v_z\frac{m}{2}\,\mathbf{v}^2\,f.
\label{eq:5.2}
\end{equation}
To find the relaxation equation of the energy current, we assume that the system is in mechanical equilibrium with a stream velocity $\mathbf{u}=0$, so that $\mathbf{U}=\mathbf{v}$ in Eq.~\ref{eq:1.4}, and we include a temperature gradient $T(\mathbf{r})$. Eq.~\ref{eq:5.2}  shows that $J_E$ vanishes for $f=f_0$, since the integrand would be odd in $v_z$.

Using Eq.~\ref{eq:5.2} and Eq.~\ref{eq:1.1} with $\tau\equiv\tau_\kappa$, we have
\begin{equation}
\dot{J}_E+\frac{m}{2}\,\partial_z\!\!\int d^3\mathbf{v}\,v_z^2\mathbf{v}^2\,f_0 =-\frac{1}{\tau_{\kappa}}\,J_E
\label{eq:5.3}
\end{equation}
or
\begin{equation}
\dot{J}_E +\frac{1}{\tau_{\kappa}}\,J_E =-\partial_z I_{zz}.
\label{eq:5.4}
\end{equation}
Here
\begin{equation}
I_{zz} \equiv \frac{m}{2}\,\int d^3\mathbf{v}\,v_z^2\,\mathbf{v}^2\,f_0.
\label{eq:5.5}
\end{equation}

To evaluate Eq.~\ref{eq:5.5}, we require an explicit form for the equilibrium phase space distribution, which we take to be a Maxwell-Boltzmann distribution,
$f_0(\mathbf{r},\mathbf{v}) = n(\mathbf{r})\,W_0(\mathbf{v})$, where
\begin{equation}
W_0(\mathbf{v})=\frac{e^{-\mathbf{v}^2/v_0^2}}{(v_0\sqrt{\pi})^3}
\label{eq:5.7}
\end{equation}
with $v_0 = \sqrt{2k_BT/m}$ the thermal speed.  As discussed above, we assume that $T=T(\mathbf{r})$ spatially varies, producing a temperature gradient, as discussed further below. Then,
\begin{equation}
I_{zz} =\frac{m}{2}\,\frac{1}{3}\,\langle v^4\rangle\, n(\mathbf{r}).
\label{eq:5.8}
\end{equation}
Here,
\begin{equation}
 \langle v^4\rangle  = \int d^3\mathbf{v}\,v^4\,W_0(\mathbf{v})=\frac{4}{\sqrt{\pi}}v_0^4\int d\left(\frac{v}{v_0}\right)\left(\frac{v}{v_0}\right)^6\,e^{-(v/v_0)^2},
 \label{eq:5.10}
\end{equation}
which yields
\begin{equation}
 \langle v^4\rangle = \frac{15}{4}\,v_0^4 = 15\,\left(\frac{k_BT}{m}\right)^2.
 \label{eq:6.3}
\end{equation}
Using Eq.~\ref{eq:6.3} in Eq.~\ref{eq:5.8}, and the pressure  $p(\mathbf{r})=n(\mathbf{r})\,k_BT(\mathbf{r})$ from Eqs.~\ref{eq:1.8}~and~\ref{eq:5.7}, we have
\begin{equation}
I_{zz} = \frac{5}{2}\frac{k_B}{m}\,n(\mathbf{r})\,k_B T^2(\mathbf{r})=\frac{5}{2}\frac{k_B}{m}\,p(\mathbf{r})\,T(\mathbf{r}).
\label{eq:6.7}
\end{equation}
With Eq.~\ref{eq:5.4}, we then obtain
\begin{equation}
\dot{J}_E+\frac{1}{\tau_\kappa}\,J_E = -\frac{5}{2}\frac{k_B}{m}\,\partial_z[\,p(\mathbf{r})\,T(\mathbf{r})\,].
\label{eq:6.8}
\end{equation}
For pure heat flow, mechanical equilibrium requires
\begin{equation}
\nabla p(\mathbf{r}) = 0.
\label{eq:6.9}
\end{equation}
Therefore, Eq.~\ref{eq:6.8} becomes
\begin{equation}
\dot{J}_E+\frac{1}{\tau_\kappa}J_E = -\frac{5}{2}\frac{k_B}{m}\,p\,\,\partial_z\delta T,
\label{eq:6.10}
\end{equation}
where we suppress the argument $\mathbf{r}$. Here, the temperature $T=T_0+\delta T$, with $T_0$ the uniform background temperature, so that $\partial_z T=\partial_z\delta T$.

For fast relaxation, where $\dot{J}_E\ll J_E/\tau_{\kappa}$, Eq.~\ref{eq:6.10} gives
\begin{equation}
J_E\rightarrow -\frac{5}{2}\frac{k_B}{m}\,\tau_{\kappa}\,p\,\,\partial_z\delta T \equiv -\kappa_T\,\partial_z\delta T,
\label{eq:7.1}
\end{equation}
where the static thermal conductivity is
\begin{equation}
\kappa_T= \frac{5}{2}\frac{k_B}{m}\,\tau_{\kappa}\,p\,.
\label{eq:7.2}
\end{equation}

We find $\delta\dot{Q}_\kappa$ using Eq.~\ref{eq:deltaQkappa}, $\delta Q_\kappa\equiv -(\beta/c_{V_1})\,\partial_z\, J_E$. Then, operating on Eq.~\ref{eq:6.10} with $-(\beta/c_{V_1})\,\partial_z$, we have
\begin{equation}
\delta\dot{Q}_\kappa+\frac{1}{\tau_\kappa}\delta{Q}_\kappa = \frac{5}{2}\frac{k_B}{m}\,\frac{p}{n_0c_{V_1}}\,\,\partial^2_z\,\delta\tilde{T},
\label{eq:8.2}
\end{equation}
where $\delta\tilde{T} = n_0\beta\,\delta T$ (see Eq.~\ref{eq:2.5H}).
For fast relaxation, where $\delta\dot{Q}_\kappa\ll\frac{1}{\tau_\kappa}\delta{Q}_\kappa$,
Eq.~\ref{eq:8.2} gives
\begin{equation}
\delta{Q}_\kappa\rightarrow \frac{5}{2}\frac{k_B}{m}\,\frac{\tau_\kappa\,p}{n_0c_{V_1}}\,\partial^2_z\delta\tilde{T}
= \frac{\kappa_T}{n_0c_{V_1}}\,\partial^2_z\,\delta\tilde{T}.
\label{eq:8.3}
\end{equation}
In this limit, Eq.~\ref{eq:1.9Sd} with Eq.~\ref{eq:8.3} reproduces the Navier-Stokes form used in Ref.~\cite{XinHydroRelax}.

We note that the coefficient $5/2$ in Eq.~\ref{eq:7.2} for $\kappa_T$ and in Eq.~\ref{eq:8.2} for $\delta\dot{Q}_\kappa$ is dependent on the assumed Maxwell-Boltzmann approximation for the phase space distribution $f_0$, which was needed to evaluate Eq.~\ref{eq:5.5}, yielding Eq.~\ref{eq:6.7}. However, in calculations of the transport times for a unitary Fermi gas, it has been found that Pauli blocking appears to be nearly cancelled by in-medium scattering. This yields transport times that are remarkably close to those obtained using a Boltzmann approximation~\cite{EnssTransport}.

\subsection{Relaxation of the Fourier Components}
In the experiments, as described in the main text, a spatially periodic perturbing potential $\delta U(z)$ is used to create a spatially periodic density perturbation $\delta n(z,t=0)$, with a wavelength $\lambda$ and corresponding wavevector $q=2\pi/\lambda$. After the system reaches  mechanical and thermal equilibrium, $\delta U$ is abruptly extinguished and $\delta n(z,t)$
is measured. A spatial Fourier transform of the relaxing density perturbation $\delta n(z,t)$ yields the time dependence of the Fourier component, $\delta n(q,t)$.
The evolution of $\delta n(q,t)$ is readily determined from the Fourier transforms of Eqs.~\ref{eq:2.5bS},~\ref{eq:4.7},~\ref{eq:1.9Sd}~and~\ref{eq:8.2}. Defining $\delta n(q,t)\equiv \delta n$, $\delta\tilde{T}(q,t)\equiv\delta\tilde{T}$, $\delta Q_\eta(q,t)\equiv\delta Q_\eta$, and $\delta Q_\kappa(q,t)\equiv\delta Q_\kappa$, we find
\begin{eqnarray}
\delta\ddot{n}=-\omega_T^2\,\delta n-\omega_T^2\,\delta\tilde{T}+\delta Q_\eta\hspace{0.25in}{\rm with}\hspace{0.25in}\omega_T\equiv c_Tq\,,
\label{eq:1.5FT}
\end{eqnarray}
\begin{equation}
\delta\dot{Q_{\eta}}=-\frac{1}{\tau_{\eta}}\delta Q_{\eta}-\Omega_\eta^2\,\delta\dot{n}\hspace{0.25in}{\rm with}\hspace{0.25in}\Omega_\eta^2\equiv\frac{4}{3}\,\frac{p}{mn_0}\,q^2\,,
\label{eq:4.3FT}
\end{equation}
and
\begin{equation}
\delta\dot{\tilde{T}}=\epsilon_{LP}\,\delta\dot{n}+\delta Q_\kappa\,,\hspace{0.25in}{\rm where}
\label{eq:1.7FT}
\end{equation}
\begin{equation}
\delta\dot{Q}_\kappa=-\frac{1}{\tau_\kappa}\delta{Q}_\kappa -\Omega_\kappa^2\,\delta\tilde{T}\hspace{0.25in}{\rm with}\hspace{0.25in}\Omega_\kappa^2=\frac{5}{2}\frac{k_B}{m}\,\frac{p}{n_0 c_{V_1}}\,\,q^2.
\label{eq:8.2FT}
\end{equation}

For a unitary Fermi gas, the pressure $p$ and internal energy density ${\cal E}$ have the universal forms
\begin{equation}
p=\frac{2}{5}\,n\epsilon_F(n)\,f_E(\theta)=\frac{2}{3}\,{\cal E},
\label{eq:1.9FT}
\end{equation}
where the universal function $f_E(\theta)$ has been measured~\cite{KuThermo} as a function of the reduced temperature $\theta\equiv T/T_F$. Here, the local Fermi energy $\epsilon_F(n)=\frac{\hbar^2}{2m}(3\pi^2 n)^{2/3}$ with $n$ the total density for a 50-50 mixture of two spin states. Taking  $\epsilon_F(n)= mv_F^2/2$, which defines the Fermi speed $v_F$, we can write in Eq.~\ref{eq:4.3FT}
\begin{equation}
\Omega_\eta^2\equiv\frac{4}{15}\,\omega_F^2\,f_E(\theta) \hspace{0.25in}{\rm with}\hspace{0.25in} \omega_F\equiv v_Fq\,.
\label{eq:1.11FT}
\end{equation}
Similarly, we can write in Eq.~\ref{eq:8.2FT}
\begin{equation}
\Omega_\kappa^2\equiv\frac{1}{2}\,\frac{k_B}{c_{V_1}}\,\omega_F^2\,f_E(\theta).
\label{eq:1.12FT}
\end{equation}
In the short relaxation time limit, we see that $\delta Q_\eta\rightarrow -\gamma_\eta\delta\dot{n}$, with $\gamma_\eta\equiv\tau_\eta\Omega_\eta^2=\frac{4}{3}\frac{\eta}{n_0m}\,q^2$ and $\delta Q_\kappa\rightarrow -\gamma_\kappa\delta\tilde{T}$, with $\gamma_\kappa\equiv\tau_\kappa\Omega_\kappa^2=\frac{\kappa_T}{n_0c_{V_1}}\,q^2$, reproducing the results in the supplement of Ref.~\cite{XinHydroRelax}.

\subsection{Data Analysis Method}

\begin{figure}[htb]
\centering
\includegraphics[width=3.0in]{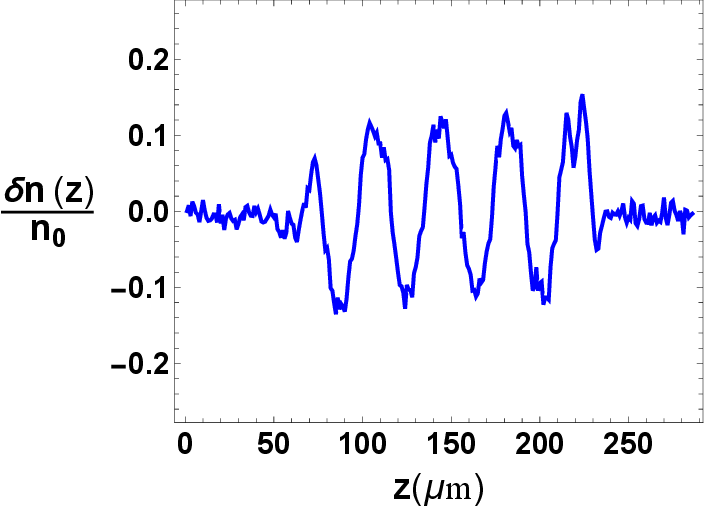}
\caption{Initial density perturbation for a wavelength $\lambda = 37.0\,\mu$m.  \label{fig:IC}}
\end{figure}

Initially, a spatially periodic repulsive potential is applied to the sample, creating a spatially periodic density profile in static equilibrium, Fig.~\ref{fig:IC}. A linear response is obtained by using a small amplitude, typically  $\pm 10$\%  of the average central density $n_0$.  We observe a negligible variation in the measured properties, within our error bars, for data obtained using amplitudes between between $\pm 10$\% and  $\pm 15$\%.
After the perturbation is extinguished, the evolution of the dominant Fourier component $\delta n(q,t)$ is observed as a function of time and analyzed as discussed below.

\begin{figure}[htb]
\centering
\includegraphics[width=3.0in]{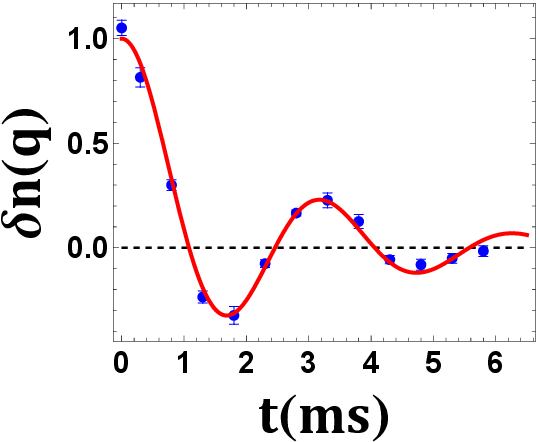}
\caption{Relaxation model fit for $T/T_F=0.37$ and $\lambda = 37.0\,\mu$m. Data (blue dots), model (red).  \label{fig:dataFit}}
\end{figure}

Analytic solutions to the linear time-dependent equations~\ref{eq:1.5FT},~\ref{eq:4.3FT},~\ref{eq:1.7FT},~and~\ref{eq:8.2FT} are easily obtained for a given set of fit parameters, with the initial conditions $\delta n(q,0)=A$, $\delta\tilde{T}(q,0)=0$, $\delta Q_\eta(q,0)=0$, and $\delta Q_\kappa(q,0)=0$. As described in more detail below, the solution for $\delta n(q,t)$ is fit to the data using the amplitude $A$, isothermal frequency $\omega_T$, and relaxation times $\tau_\eta$ and $\tau_\kappa$ as free parameters. As in our previous experiments~\cite{XinHydroRelax}, the ratio $\omega_T/\omega_F$ self consistently determines the reduced temperature $\theta$ in the fits. A typical fit to data taken for $T/T_F=0.37$ with $\lambda = 37.0\,\mu$m is shown in Fig.~\ref{fig:dataFit}. Here, we have divided the data and the predicted $\delta n(q,t)$ by the fit amplitude $A$, so that the peak density response is normalized to unity, $\delta n(q,0)=1$.

In the analysis of the data, we employ a $\chi^2$ fit, holding the amplitude $A$ constant, to determine the three parameters $\omega_T$, $\tau_\eta$ and $\tau_\kappa$ by minimizing $\chi^2$. Then $A$ is changed, and the three parameters are again determined to give a new $\chi^2$. Repeating this process determines the global minimum for all four parameters, with a typical $\chi^2$ per degree of freedom $\simeq 1.5$.

We note that the number of fit parameters (4) is the same as for the fast relaxation approximation used in our previous work~\cite{XinHydroRelax}, where $A$, $\omega_T$, $\eta$, and $\kappa$ were the fit parameters.
In the short transport time limit, $\tau_\eta$ and $\tau_\kappa$ are simply caveats for $\eta$ and $\kappa_T$, as can be seen from Eqs.~\ref{eq:4.3}~and~\ref{eq:7.2}. In this case, we recover the Navier-Stokes description, where the extracted static values of $\eta$ and $\kappa_T$ are independent of the time-dependent kinetic theory model.  However, by explicitly including the time dependence of the viscous force and the heat current, which are driven at several frequencies in the free decay, the new model consistently corrects the extracted static transport properties for the small but finite response times, shifting the extracted static transport properties upward relative to their finite frequency values.

\begin{figure}[htb]
\centering
\includegraphics[width=3.0in]{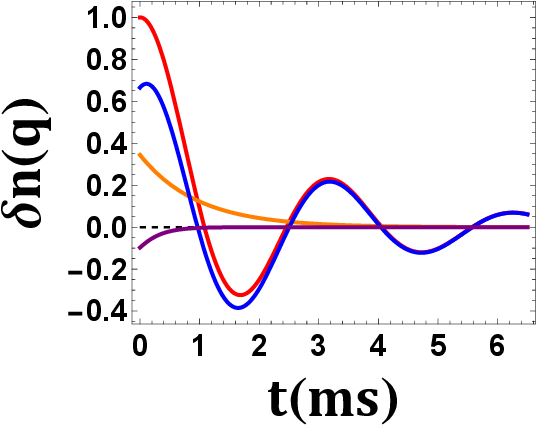}\hspace*{0.25in}\includegraphics[width=3.0in]{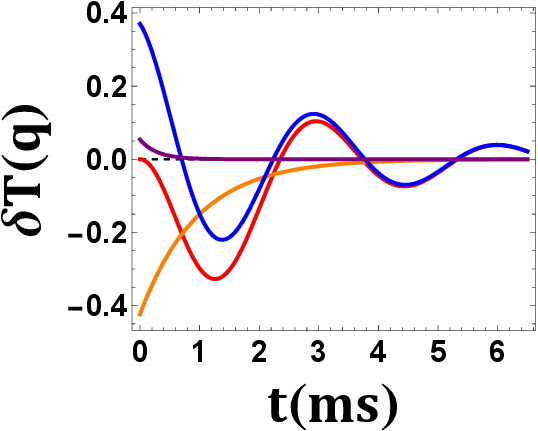}
\caption{Density and temperature responses (red) for $T/T_F=0.37$ and $\lambda = 37.0\,\mu$m showing the mode content.  First sound  mode (blue), thermally diffusive mode (orange), fast relaxation modes  (purple). Note that the fast relaxation modes are increased $\times 10$ in the density response to make them visible, $\times\,1$ in temperature response.  \label{fig:densitytempmodes}}
\end{figure}

From the fit, we obtain analytic solutions for all of the responses $\delta n(q,t)$, $\delta\tilde{T}(q,t)$, $\delta Q_\eta(q,t)$, and $\delta Q_\kappa(q,t)$. For each response, the analytic solution determines the contributions of 4 different modes, as shown in Fig.~\ref{fig:densitytempmodes} for the density and temperature perturbations.
Two fast modes (purple) arise predominantly from relaxation of the viscous force and heat current to their Navier-Stokes forms on time scales of $\tau_\eta$ and $\tau_\kappa$.
A zero-frequency thermally-diffusive mode (orange) decays exponentially at a rate determined by $\tau_\kappa$, which controls the effective frequency dependent thermal conductivity. Finally, an oscillating first sound mode (blue) decays at a rate determined by $\tau_\eta$ and $\tau_\kappa$, which control the effective frequency dependent sound diffusivity.

For the density perturbation, the fast modes contribute only  1\%  at $t=0$, and are multiplied by a factor of $10$ in Fig.~\ref{fig:densitytempmodes} to make them visible. Their contribution is not directly observed in the experiments. In contrast, the thermally diffusive mode contributes $\simeq 35$\% to the density perturbation, while the first sound mode contributes $\simeq 65$\%. For this reason, the fit parameters, comprising the frequency $\omega_T$ and the transport times $\tau_\eta$ and $\tau_\kappa$, which determine the oscillation frequency and decay rates of the two dominant modes, are nearly independent, simplifying the fit procedure.

\begin{figure}[htb]
\centering
\includegraphics[width=3.0in]{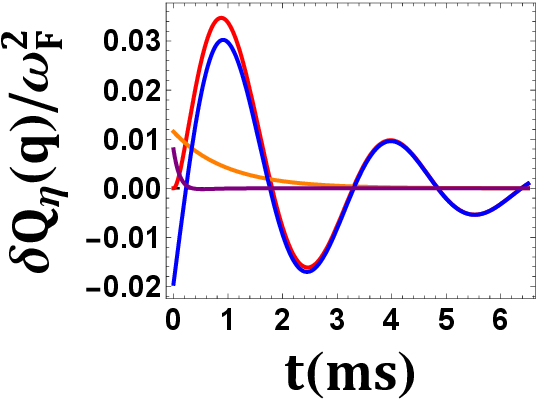}\hspace*{0.25in}\includegraphics[width=3.0in]{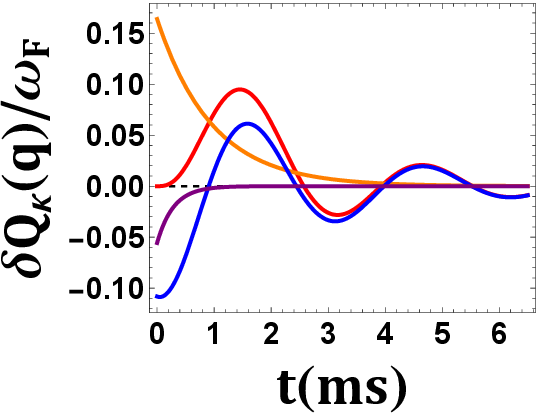}
\caption{Viscous force and heat current responses (red) for $T/T_F=0.37$ and $\lambda = 37.0\,\mu$m showing the mode content. First sound  mode (blue), thermally diffusive mode (orange), fast relaxation modes $\times\,1$ (purple).  \label{fig:modes2}}
\end{figure}

In contrast to the density response, the fast modes contribute significantly $\simeq 10$\%  to the short time response of the temperature perturbation, shown $\times\,1$ in Fig.~\ref{fig:densitytempmodes} and more significantly $\simeq 30$\% to the short time responses of the viscous force and heat current, shown $\times\,1$ in Fig.~\ref{fig:modes2}. We see from Eq.~\ref{eq:1.5FT} that a natural scale for $\delta Q_\eta$ is $\delta\ddot{n}(q,t)$,  where $\delta n(q,0)=1$. For this reason, we display $\delta Q_\eta$ in units of $\omega_F^2\delta n(q,0)=\omega_F^2$. Similarly, Eq.~\ref{eq:1.7FT} shows that a natural scale for $\delta Q_\kappa$ is $\delta\dot{n}(q,t)$. In this case, we display $\delta Q_\kappa$ in units of $\omega_F\delta n(q,0)=\omega_F$.

\begin{figure}[htb]
\centering
\includegraphics[width=3.0in]{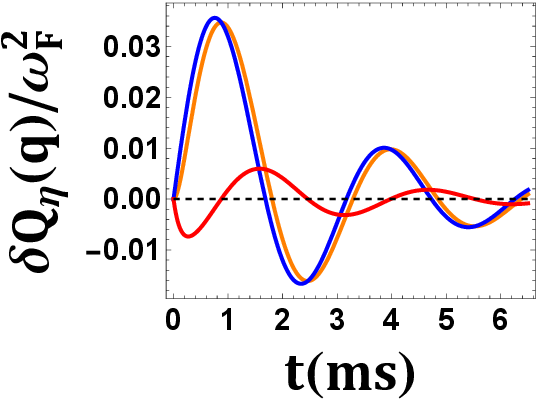}\hspace*{0.25in}\includegraphics[width=3.0in]{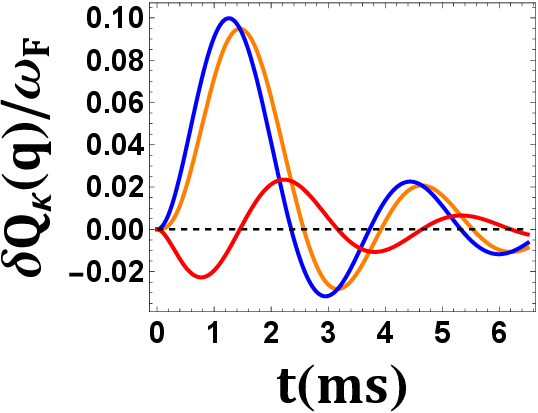}
\caption{Difference (red) between kinetic theory model (orange)  and the fast relaxation approximation (blue) for the viscous force (left) and heat current (right) with $T/T_F=0.37$ and $\lambda = 37.0\,\mu$m.   \label{fig:modes3}}
\end{figure}

Fig.~\ref{fig:modes3} shows the difference (red) between the kinetic theory relaxation model (orange) and the fast relaxation approximation (blue). We see that the deviation from the fast relaxation (long-wavelength) approximation is significant, emphasizing the need for the analysis based on the relaxation model. With the new kinetic theory model, the finite relaxation time corrections to the transport properties are consistently included in the analysis, enabling determination of the static transport properties, as discussed in \S~\ref{sec:static}.

\subsection{Static Transport Properties}
\label{sec:static}

In the main text, the relaxation times are given in units of the ``Fermi time," which we take to be $\tau_F\equiv\lambda_F/v_F$, i.e., the time for an atom to move a Fermi wavelength at the Fermi speed.
With $\lambda_F=2\pi\hbar/(m v_F)$, and $m v_F^2=2\epsilon_F$, the Fermi time is
\begin{equation}
\tau_F=\frac{\pi\hbar}{\epsilon_F}.
\label{eq:4.5FT}
\end{equation}

With the relaxation times determined from the fits, the {\it static} shear viscosity Eq.~\ref{eq:4.3} is determined in units of $\hbar n_0$ by
\begin{equation}
\eta=\tau_\eta\,p=\alpha_\eta\,\hbar n_0\, ,
\label{eq:3.5FT}
\end{equation}
where  Eq.~\ref{eq:1.9FT} for $p$ gives the dimensionless shear viscosity coefficient,
\begin{equation}
\alpha_\eta=\frac{2\pi}{5}\,\frac{\tau_\eta}{\tau_F}\,f_E(\theta).
\label{eq:3.6FT}
\end{equation}

Similarly, the {\it static} thermal conductivity Eq.~\ref{eq:7.2} is determined in units of $\hbar n_0\,k_B/m$ by
\begin{equation}
\kappa_T=\frac{5}{2}\frac{k_B}{m}\,\tau_\kappa\,p=\alpha_\kappa\,\frac{k_B}{m}\,\hbar n_0\, ,
\label{eq:3.7FT}
\end{equation}
where the dimensionless thermal conductivity coefficient is
\begin{equation}
\alpha_\kappa=\pi\,\frac{\tau_\kappa}{\tau_F}\,f_E(\theta).
\label{eq:3.8FT}
\end{equation}

Finally, the first sound diffusivity~\cite{XinHydroRelax,LandauFluids} is determined from the static transport properties,
\begin{equation}
D_1=\frac{4}{3}\frac{\eta}{n_0m}+\left(\frac{1}{c_{V_1}}-\frac{1}{c_{P_1}}\right)\frac{\kappa_T}{n_0}.
\label{eq:4.1D1}
\end{equation}
For a unitary Fermi gas in a 50-50 mixture of two spin states and total density $n$~\cite{XinHydroRelax,MZSound},
\begin{equation}
\frac{1}{c_{V_1}}-\frac{1}{c_{P_1}}=\frac{4}{15}\frac{nT}{p}=\frac{1}{k_B}\frac{2}{3}\frac{\theta}{f_E(\theta)}.
\label{eq:4.2D1}
\end{equation}
Then
\begin{equation}
\frac{D_1}{\hbar/m}=\frac{4}{3}\,\alpha_\eta+\frac{2}{3}\frac{\theta}{f_E(\theta)}\,\alpha_\kappa.
\label{eq:4.3D1}
\end{equation}
Using Eqs.~\ref{eq:3.6FT}~and~\ref{eq:3.8FT}, we then obtain $D_1$ in units of $\hbar/m$ from the measured relaxation times,
\begin{equation}
\frac{D_1}{\hbar/m}=\frac{8\pi}{15}\,\frac{\tau_\eta}{\tau_F}\,f_E(\theta)+
\frac{2\pi}{3}\,\frac{\tau_\kappa}{\tau_F}\,\theta.
\label{eq:4.4D1}
\end{equation}

For comparison with the static transport properties shown in the main text, which are obtained from the measured relaxation times using Eqs.~\ref{eq:3.6FT},~\ref{eq:3.8FT},~and~\ref{eq:4.4D1}, we parameterize $\alpha_\eta$ as
\begin{equation}
\alpha_\eta=\alpha_{3/2}\,\theta^{3/2}+\alpha_{2\eta}.
\label{eq:5.3Analytic}
\end{equation}
Here, the first term is the high temperature limit, which is obtained from a variational calculation~\cite{BruunViscousNormalDamping},  $\alpha_{3/2}=45\pi^{3/2}/(64\sqrt{2})=2.76849\simeq 2.77$ and $\alpha_{2\eta}$ is the universal density shift coefficient, which is used as a fit parameter. Similarly, we parameterize $\alpha_\kappa$ as
\begin{equation}
\alpha_\kappa=\frac{15}{4}\,\left(\alpha_{3/2}\,\theta^{3/2}+\alpha_{2\kappa}\right).
\label{eq:6.3Analytic}
\end{equation}
Here, the leading factor of $15/4$ is chosen to yield the correct high temperature limit for the ratio $\kappa_T/\eta$. In this limit, Eqs.~\ref{eq:3.7FT}~and~\ref{eq:3.5FT} yield  $\kappa_T=\frac{15}{4}\frac{k_B}{m}\,\eta$, where we have used $\tau_\kappa/\tau_\eta=3/2$, which can be shown to hold for any isotropic collision cross section.

With Eqs.~\ref{eq:5.3Analytic}~and~\ref{eq:6.3Analytic}, the relaxation times are then parameterized in the main text by inverting
Eqs.~\ref{eq:3.6FT}~and~\ref{eq:3.8FT} as
\begin{equation}
\label{eq:6.4Analytic}
\frac{\tau_\eta}{\tau_F}=\alpha_\eta\,\frac{5}{2\pi}\frac{1}{f_E(\theta)}
\end{equation}
and
\begin{equation}
\label{eq:6.5Analytic}
\frac{\tau_\kappa}{\tau_F}=\alpha_\kappa\,\frac{1}{\pi}\frac{1}{f_E(\theta)}.
\end{equation}


\begin{thebibliography}{35}%
\makeatletter
\providecommand \@ifxundefined [1]{%
 \@ifx{#1\undefined}
}%
\providecommand \@ifnum [1]{%
 \ifnum #1\expandafter \@firstoftwo
 \else \expandafter \@secondoftwo
 \fi
}%
\providecommand \@ifx [1]{%
 \ifx #1\expandafter \@firstoftwo
 \else \expandafter \@secondoftwo
 \fi
}%
\providecommand \natexlab [1]{#1}%
\providecommand \enquote  [1]{``#1''}%
\providecommand \bibnamefont  [1]{#1}%
\providecommand \bibfnamefont [1]{#1}%
\providecommand \citenamefont [1]{#1}%
\providecommand \href@noop [0]{\@secondoftwo}%
\providecommand \href [0]{\begingroup \@sanitize@url \@href}%
\providecommand \@href[1]{\@@startlink{#1}\@@href}%
\providecommand \@@href[1]{\endgroup#1\@@endlink}%
\providecommand \@sanitize@url [0]{\catcode `\\12\catcode `\$12\catcode
  `\&12\catcode `\#12\catcode `\^12\catcode `\_12\catcode `\%12\relax}%
\providecommand \@@startlink[1]{}%
\providecommand \@@endlink[0]{}%
\providecommand \url  [0]{\begingroup\@sanitize@url \@url }%
\providecommand \@url [1]{\endgroup\@href {#1}{\urlprefix }}%
\providecommand \urlprefix  [0]{URL }%
\providecommand \Eprint [0]{\href }%
\providecommand \doibase [0]{https://doi.org/}%
\providecommand \selectlanguage [0]{\@gobble}%
\providecommand \bibinfo  [0]{\@secondoftwo}%
\providecommand \bibfield  [0]{\@secondoftwo}%
\providecommand \translation [1]{[#1]}%
\providecommand \BibitemOpen [0]{}%
\providecommand \bibitemStop [0]{}%
\providecommand \bibitemNoStop [0]{.\EOS\space}%
\providecommand \EOS [0]{\spacefactor3000\relax}%
\providecommand \BibitemShut  [1]{\csname bibitem#1\endcsname}%
\let\auto@bib@innerbib\@empty
\bibitem [{\citenamefont {Adams}\ \emph {et~al.}(2012)\citenamefont {Adams},
  \citenamefont {Carr}, \citenamefont {{Sch\"{a}fer}}, \citenamefont
  {Steinberg},\ and\ \citenamefont {Thomas}}]{NJPReview}%
  \BibitemOpen
  \bibfield  {author} {\bibinfo {author} {\bibfnamefont {A.}~\bibnamefont
  {Adams}}, \bibinfo {author} {\bibfnamefont {L.~D.}\ \bibnamefont {Carr}},
  \bibinfo {author} {\bibfnamefont {T.}~\bibnamefont {{Sch\"{a}fer}}}, \bibinfo
  {author} {\bibfnamefont {P.}~\bibnamefont {Steinberg}},\ and\ \bibinfo
  {author} {\bibfnamefont {J.~E.}\ \bibnamefont {Thomas}},\ }\bibfield  {title}
  {\bibinfo {title} {Strongly correlated quantum fluids: ultracold quantum
  gases, quantum chromodynamic plasmas and holographic duality},\ }\href@noop
  {} {\bibfield  {journal} {\bibinfo  {journal} {New J. Phys.}\ }\textbf
  {\bibinfo {volume} {14}},\ \bibinfo {pages} {115009} (\bibinfo {year}
  {2012})}\BibitemShut {NoStop}%
\bibitem [{\citenamefont {Bloch}\ \emph {et~al.}(2012)\citenamefont {Bloch},
  \citenamefont {Dalibard},\ and\ \citenamefont {Nascimb\`ene}}]{BlochReview}%
  \BibitemOpen
  \bibfield  {author} {\bibinfo {author} {\bibfnamefont {I.}~\bibnamefont
  {Bloch}}, \bibinfo {author} {\bibfnamefont {J.}~\bibnamefont {Dalibard}},\
  and\ \bibinfo {author} {\bibfnamefont {S.}~\bibnamefont {Nascimb\`ene}},\
  }\bibfield  {title} {\bibinfo {title} {Quantum simulations with ultracold
  quantum gases},\ }\href {https://doi.org/10.1038/nphys2259} {\bibfield
  {journal} {\bibinfo  {journal} {Nature Physics}\ }\textbf {\bibinfo {volume}
  {8}},\ \bibinfo {pages} {267} (\bibinfo {year} {2012})}\BibitemShut {NoStop}%
\bibitem [{\citenamefont {Strinati}\ \emph {et~al.}(2018)\citenamefont
  {Strinati}, \citenamefont {Pieri}, \citenamefont {R\"{o}pke}, \citenamefont
  {Schuck},\ and\ \citenamefont {Urban}}]{UrbanReview}%
  \BibitemOpen
  \bibfield  {author} {\bibinfo {author} {\bibfnamefont {G.~C.}\ \bibnamefont
  {Strinati}}, \bibinfo {author} {\bibfnamefont {P.}~\bibnamefont {Pieri}},
  \bibinfo {author} {\bibfnamefont {G.}~\bibnamefont {R\"{o}pke}}, \bibinfo
  {author} {\bibfnamefont {P.}~\bibnamefont {Schuck}},\ and\ \bibinfo {author}
  {\bibfnamefont {M.}~\bibnamefont {Urban}},\ }\bibfield  {title} {\bibinfo
  {title} {The {BCS-BEC} crossover: From ultra-cold {Fermi} gases to nuclear
  systems},\ }\href@noop {} {\bibfield  {journal} {\bibinfo  {journal} {Physics
  Reports}\ }\textbf {\bibinfo {volume} {738}},\ \bibinfo {pages} {1} (\bibinfo
  {year} {2018})}\BibitemShut {NoStop}%
\bibitem [{\citenamefont {Bruun}\ and\ \citenamefont
  {Smith}(2007)}]{BruunViscousNormalDamping}%
  \BibitemOpen
  \bibfield  {author} {\bibinfo {author} {\bibfnamefont {G.~M.}\ \bibnamefont
  {Bruun}}\ and\ \bibinfo {author} {\bibfnamefont {H.}~\bibnamefont {Smith}},\
  }\bibfield  {title} {\bibinfo {title} {Shear viscosity and damping for a
  {Fermi} gas in the unitary limit},\ }\href@noop {} {\bibfield  {journal}
  {\bibinfo  {journal} {Phys. Rev. A}\ }\textbf {\bibinfo {volume} {75}},\
  \bibinfo {pages} {043612} (\bibinfo {year} {2007})}\BibitemShut {NoStop}%
\bibitem [{\citenamefont {Taylor}\ and\ \citenamefont
  {Randeria}(2010)}]{RanderiaViscosity}%
  \BibitemOpen
  \bibfield  {author} {\bibinfo {author} {\bibfnamefont {E.}~\bibnamefont
  {Taylor}}\ and\ \bibinfo {author} {\bibfnamefont {M.}~\bibnamefont
  {Randeria}},\ }\bibfield  {title} {\bibinfo {title} {Viscosity of strongly
  interacting quantum fluids: Spectral functions and sum rules},\ }\href@noop
  {} {\bibfield  {journal} {\bibinfo  {journal} {Phys. Rev. A}\ }\textbf
  {\bibinfo {volume} {81}},\ \bibinfo {pages} {053610} (\bibinfo {year}
  {2010})}\BibitemShut {NoStop}%
\bibitem [{\citenamefont {Braby}\ \emph {et~al.}(2010)\citenamefont {Braby},
  \citenamefont {Chao},\ and\ \citenamefont
  {Sch\"afer}}]{BrabySchaeferThermalCond}%
  \BibitemOpen
  \bibfield  {author} {\bibinfo {author} {\bibfnamefont {M.}~\bibnamefont
  {Braby}}, \bibinfo {author} {\bibfnamefont {J.}~\bibnamefont {Chao}},\ and\
  \bibinfo {author} {\bibfnamefont {T.}~\bibnamefont {Sch\"afer}},\ }\bibfield
  {title} {\bibinfo {title} {Thermal conductivity and sound attenuation in
  dilute atomic {Fermi} gases},\ }\href@noop {} {\bibfield  {journal} {\bibinfo
   {journal} {Phys. Rev. A}\ }\textbf {\bibinfo {volume} {82}},\ \bibinfo
  {pages} {033619} (\bibinfo {year} {2010})}\BibitemShut {NoStop}%
\bibitem [{\citenamefont {Braby}\ \emph {et~al.}(2011)\citenamefont {Braby},
  \citenamefont {Chao},\ and\ \citenamefont {Schäfer}}]{DrudeBraby2011}%
  \BibitemOpen
  \bibfield  {author} {\bibinfo {author} {\bibfnamefont {M.}~\bibnamefont
  {Braby}}, \bibinfo {author} {\bibfnamefont {J.}~\bibnamefont {Chao}},\ and\
  \bibinfo {author} {\bibfnamefont {T.}~\bibnamefont {Schäfer}},\ }\bibfield
  {title} {\bibinfo {title} {Viscosity spectral functions of the dilute {Fermi}
  gas in kinetic theory},\ }\href@noop {} {\bibfield  {journal} {\bibinfo
  {journal} {New Journal of Physics}\ }\textbf {\bibinfo {volume} {13}},\
  \bibinfo {pages} {035014} (\bibinfo {year} {2011})}\BibitemShut {NoStop}%
\bibitem [{\citenamefont {Enss}\ \emph {et~al.}(2011)\citenamefont {Enss},
  \citenamefont {Haussmann},\ and\ \citenamefont {Zwerger}}]{ZwergerViscosity}%
  \BibitemOpen
  \bibfield  {author} {\bibinfo {author} {\bibfnamefont {T.}~\bibnamefont
  {Enss}}, \bibinfo {author} {\bibfnamefont {R.}~\bibnamefont {Haussmann}},\
  and\ \bibinfo {author} {\bibfnamefont {W.}~\bibnamefont {Zwerger}},\
  }\bibfield  {title} {\bibinfo {title} {Viscosity and scale invariance in the
  unitary {Fermi} gas},\ }\href@noop {} {\bibfield  {journal} {\bibinfo
  {journal} {Annals Phys.}\ }\textbf {\bibinfo {volume} {326}},\ \bibinfo
  {pages} {770} (\bibinfo {year} {2011})}\BibitemShut {NoStop}%
\bibitem [{\citenamefont {Guo}\ \emph {et~al.}(2011)\citenamefont {Guo},
  \citenamefont {Wulin}, \citenamefont {Chien},\ and\ \citenamefont
  {Levin}}]{LevinViscosity}%
  \BibitemOpen
  \bibfield  {author} {\bibinfo {author} {\bibfnamefont {H.}~\bibnamefont
  {Guo}}, \bibinfo {author} {\bibfnamefont {D.}~\bibnamefont {Wulin}}, \bibinfo
  {author} {\bibfnamefont {C.-C.}\ \bibnamefont {Chien}},\ and\ \bibinfo
  {author} {\bibfnamefont {K.}~\bibnamefont {Levin}},\ }\bibfield  {title}
  {\bibinfo {title} {Microscopic approach to shear viscosities of uunitary
  {Fermi} gases above and below the superfluid transition},\ }\href@noop {}
  {\bibfield  {journal} {\bibinfo  {journal} {Phys. Rev.Lett.}\ }\textbf
  {\bibinfo {volume} {107}},\ \bibinfo {pages} {020403} (\bibinfo {year}
  {2011})}\BibitemShut {NoStop}%
\bibitem [{\citenamefont {Wlazłowski}\ \emph {et~al.}(2012)\citenamefont
  {Wlazłowski}, \citenamefont {Magierski},\ and\ \citenamefont
  {Drut}}]{DrutViscosity}%
  \BibitemOpen
  \bibfield  {author} {\bibinfo {author} {\bibfnamefont {G.}~\bibnamefont
  {Wlazłowski}}, \bibinfo {author} {\bibfnamefont {P.}~\bibnamefont
  {Magierski}},\ and\ \bibinfo {author} {\bibfnamefont {J.~E.}\ \bibnamefont
  {Drut}},\ }\bibfield  {title} {\bibinfo {title} {Shear viscosity of a unitary
  {Fermi} gas},\ }\href@noop {} {\bibfield  {journal} {\bibinfo  {journal}
  {Phys. Rev. Lett.}\ }\textbf {\bibinfo {volume} {109}},\ \bibinfo {pages}
  {020406} (\bibinfo {year} {2012})}\BibitemShut {NoStop}%
\bibitem [{\citenamefont {Romatschke}\ and\ \citenamefont
  {Young}(2013)}]{RomatschkeShear}%
  \BibitemOpen
  \bibfield  {author} {\bibinfo {author} {\bibfnamefont {P.}~\bibnamefont
  {Romatschke}}\ and\ \bibinfo {author} {\bibfnamefont {R.~E.}\ \bibnamefont
  {Young}},\ }\bibfield  {title} {\bibinfo {title} {Implications of
  hydrodynamic fluctuations for the minimum shear viscosity of the dilute
  {Fermi} gas at unitarity},\ }\href@noop {} {\bibfield  {journal} {\bibinfo
  {journal} {Phys. Rev. A}\ }\textbf {\bibinfo {volume} {87}},\ \bibinfo
  {pages} {053606} (\bibinfo {year} {2013})}\BibitemShut {NoStop}%
\bibitem [{\citenamefont {Bluhm}\ \emph {et~al.}(2017)\citenamefont {Bluhm},
  \citenamefont {Hou},\ and\ \citenamefont
  {Sch\"afer}}]{BluhmSchaeferLocalViscosity}%
  \BibitemOpen
  \bibfield  {author} {\bibinfo {author} {\bibfnamefont {M.}~\bibnamefont
  {Bluhm}}, \bibinfo {author} {\bibfnamefont {J.}~\bibnamefont {Hou}},\ and\
  \bibinfo {author} {\bibfnamefont {T.}~\bibnamefont {Sch\"afer}},\ }\bibfield
  {title} {\bibinfo {title} {Determination of the density and temperature
  dependence of the shear viscosity of a unitary {Fermi} gas based on
  hydrodynamic flow},\ }\href@noop {} {\bibfield  {journal} {\bibinfo
  {journal} {Phys. Rev. Lett.}\ }\textbf {\bibinfo {volume} {119}},\ \bibinfo
  {pages} {065302} (\bibinfo {year} {2017})}\BibitemShut {NoStop}%
\bibitem [{\citenamefont {Frank}\ \emph {et~al.}(2020)\citenamefont {Frank},
  \citenamefont {Zwerger},\ and\ \citenamefont {Enss}}]{EnssTransport}%
  \BibitemOpen
  \bibfield  {author} {\bibinfo {author} {\bibfnamefont {B.}~\bibnamefont
  {Frank}}, \bibinfo {author} {\bibfnamefont {W.}~\bibnamefont {Zwerger}},\
  and\ \bibinfo {author} {\bibfnamefont {T.}~\bibnamefont {Enss}},\ }\bibfield
  {title} {\bibinfo {title} {Quantum critical thermal transport in the unitary
  {Fermi} gas},\ }\href@noop {} {\bibfield  {journal} {\bibinfo  {journal}
  {Phys. Rev. Research}\ }\textbf {\bibinfo {volume} {2}},\ \bibinfo {pages}
  {023301} (\bibinfo {year} {2020})}\BibitemShut {NoStop}%
\bibitem [{\citenamefont {Hofmann}(2020)}]{HofmannViscosity}%
  \BibitemOpen
  \bibfield  {author} {\bibinfo {author} {\bibfnamefont {J.}~\bibnamefont
  {Hofmann}},\ }\bibfield  {title} {\bibinfo {title} {High-temperature
  expansion of the viscosity in interacting quantum gases},\ }\href@noop {}
  {\bibfield  {journal} {\bibinfo  {journal} {Phys. Rev. A}\ }\textbf {\bibinfo
  {volume} {101}},\ \bibinfo {pages} {013620} (\bibinfo {year}
  {2020})}\BibitemShut {NoStop}%
\bibitem [{\citenamefont {Zhou}\ and\ \citenamefont
  {Ma}(2021)}]{ZhouThermalCond}%
  \BibitemOpen
  \bibfield  {author} {\bibinfo {author} {\bibfnamefont {H.}~\bibnamefont
  {Zhou}}\ and\ \bibinfo {author} {\bibfnamefont {Y.}~\bibnamefont {Ma}},\
  }\bibfield  {title} {\bibinfo {title} {Thermal conductivity of an ultracold
  {Fermi} gas in the {BCS-BEC} crossover},\ }\href@noop {} {\bibfield
  {journal} {\bibinfo  {journal} {Sci. Rep.}\ }\textbf {\bibinfo {volume}
  {11}},\ \bibinfo {pages} {1228} (\bibinfo {year} {2021})}\BibitemShut
  {NoStop}%
\bibitem [{\citenamefont {O'Hara}\ \emph {et~al.}(2002)\citenamefont {O'Hara},
  \citenamefont {Hemmer}, \citenamefont {Gehm}, \citenamefont {Granade},\ and\
  \citenamefont {Thomas}}]{OHaraScience}%
  \BibitemOpen
  \bibfield  {author} {\bibinfo {author} {\bibfnamefont {K.~M.}\ \bibnamefont
  {O'Hara}}, \bibinfo {author} {\bibfnamefont {S.~L.}\ \bibnamefont {Hemmer}},
  \bibinfo {author} {\bibfnamefont {M.~E.}\ \bibnamefont {Gehm}}, \bibinfo
  {author} {\bibfnamefont {S.~R.}\ \bibnamefont {Granade}},\ and\ \bibinfo
  {author} {\bibfnamefont {J.~E.}\ \bibnamefont {Thomas}},\ }\bibfield  {title}
  {\bibinfo {title} {Observation of a strongly interacting degenerate {Fermi}
  gas of atoms},\ }\href@noop {} {\bibfield  {journal} {\bibinfo  {journal}
  {Science}\ }\textbf {\bibinfo {volume} {298}},\ \bibinfo {pages} {2179}
  (\bibinfo {year} {2002})}\BibitemShut {NoStop}%
\bibitem [{\citenamefont {Ho}(2004)}]{HoUniversalThermo}%
  \BibitemOpen
  \bibfield  {author} {\bibinfo {author} {\bibfnamefont {T.-L.}\ \bibnamefont
  {Ho}},\ }\bibfield  {title} {\bibinfo {title} {Universal thermodynamics of
  degenerate quantum gases in the unitarity limit},\ }\href@noop {} {\bibfield
  {journal} {\bibinfo  {journal} {Phys. Rev. Lett.}\ }\textbf {\bibinfo
  {volume} {92}},\ \bibinfo {pages} {090402} (\bibinfo {year}
  {2004})}\BibitemShut {NoStop}%
\bibitem [{\citenamefont {Cao}\ \emph {et~al.}(2011)\citenamefont {Cao},
  \citenamefont {Elliott}, \citenamefont {Joseph}, \citenamefont {Wu},
  \citenamefont {Petricka}, \citenamefont {{Sch\"{a}fer}},\ and\ \citenamefont
  {Thomas}}]{CaoViscosity}%
  \BibitemOpen
  \bibfield  {author} {\bibinfo {author} {\bibfnamefont {C.}~\bibnamefont
  {Cao}}, \bibinfo {author} {\bibfnamefont {E.}~\bibnamefont {Elliott}},
  \bibinfo {author} {\bibfnamefont {J.}~\bibnamefont {Joseph}}, \bibinfo
  {author} {\bibfnamefont {H.}~\bibnamefont {Wu}}, \bibinfo {author}
  {\bibfnamefont {J.}~\bibnamefont {Petricka}}, \bibinfo {author}
  {\bibfnamefont {T.}~\bibnamefont {{Sch\"{a}fer}}},\ and\ \bibinfo {author}
  {\bibfnamefont {J.~E.}\ \bibnamefont {Thomas}},\ }\bibfield  {title}
  {\bibinfo {title} {Universal quantum viscosity in a unitary {Fermi} gas},\
  }\href@noop {} {\bibfield  {journal} {\bibinfo  {journal} {Science}\ }\textbf
  {\bibinfo {volume} {331}},\ \bibinfo {pages} {58} (\bibinfo {year}
  {2011})}\BibitemShut {NoStop}%
\bibitem [{\citenamefont {Joseph}\ \emph {et~al.}(2015)\citenamefont {Joseph},
  \citenamefont {Elliott},\ and\ \citenamefont {Thomas}}]{JosephShearNearSF}%
  \BibitemOpen
  \bibfield  {author} {\bibinfo {author} {\bibfnamefont {J.~A.}\ \bibnamefont
  {Joseph}}, \bibinfo {author} {\bibfnamefont {E.}~\bibnamefont {Elliott}},\
  and\ \bibinfo {author} {\bibfnamefont {J.~E.}\ \bibnamefont {Thomas}},\
  }\bibfield  {title} {\bibinfo {title} {Shear viscosity of a unitary {Fermi}
  gas near the superfluid phase transition},\ }\href
  {https://doi.org/10.1103/PhysRevLett.115.020401} {\bibfield  {journal}
  {\bibinfo  {journal} {Phys. Rev. Lett.}\ }\textbf {\bibinfo {volume} {115}},\
  \bibinfo {pages} {020401} (\bibinfo {year} {2015})}\BibitemShut {NoStop}%
\bibitem [{\citenamefont {Navon}\ \emph {et~al.}(2021)\citenamefont {Navon},
  \citenamefont {Smith},\ and\ \citenamefont {Hadzibabic}}]{HadzibabicBox}%
  \BibitemOpen
  \bibfield  {author} {\bibinfo {author} {\bibfnamefont {N.}~\bibnamefont
  {Navon}}, \bibinfo {author} {\bibfnamefont {R.~P.}\ \bibnamefont {Smith}},\
  and\ \bibinfo {author} {\bibfnamefont {Z.}~\bibnamefont {Hadzibabic}},\
  }\bibfield  {title} {\bibinfo {title} {Quantum gases in optical boxes},\
  }\href@noop {} {\bibfield  {journal} {\bibinfo  {journal} {Nat. Phys.}\
  }\textbf {\bibinfo {volume} {17}},\ \bibinfo {pages} {1334} (\bibinfo {year}
  {2021})}\BibitemShut {NoStop}%
\bibitem [{\citenamefont {Baird}\ \emph {et~al.}(2019)\citenamefont {Baird},
  \citenamefont {Wang}, \citenamefont {Roof},\ and\ \citenamefont
  {Thomas}}]{LorinLinearHydro}%
  \BibitemOpen
  \bibfield  {author} {\bibinfo {author} {\bibfnamefont {L.}~\bibnamefont
  {Baird}}, \bibinfo {author} {\bibfnamefont {X.}~\bibnamefont {Wang}},
  \bibinfo {author} {\bibfnamefont {S.}~\bibnamefont {Roof}},\ and\ \bibinfo
  {author} {\bibfnamefont {J.~E.}\ \bibnamefont {Thomas}},\ }\bibfield  {title}
  {\bibinfo {title} {Measuring the hydrodynamic linear response of a unitary
  {Fermi} gas},\ }\href@noop {} {\bibfield  {journal} {\bibinfo  {journal}
  {Phys. Rev. Lett.}\ }\textbf {\bibinfo {volume} {123}},\ \bibinfo {pages}
  {160402} (\bibinfo {year} {2019})}\BibitemShut {NoStop}%
\bibitem [{\citenamefont {Patel}\ \emph {et~al.}(2020)\citenamefont {Patel},
  \citenamefont {Yan}, \citenamefont {Mukherjee}, \citenamefont {Fletcher},
  \citenamefont {Struck},\ and\ \citenamefont {Zwierlein}}]{MZSound}%
  \BibitemOpen
  \bibfield  {author} {\bibinfo {author} {\bibfnamefont {P.~B.}\ \bibnamefont
  {Patel}}, \bibinfo {author} {\bibfnamefont {Z.}~\bibnamefont {Yan}}, \bibinfo
  {author} {\bibfnamefont {B.}~\bibnamefont {Mukherjee}}, \bibinfo {author}
  {\bibfnamefont {R.~J.}\ \bibnamefont {Fletcher}}, \bibinfo {author}
  {\bibfnamefont {J.}~\bibnamefont {Struck}},\ and\ \bibinfo {author}
  {\bibfnamefont {M.~W.}\ \bibnamefont {Zwierlein}},\ }\bibfield  {title}
  {\bibinfo {title} {Universal sound diffusion in a strongly interacting
  {Fermi} gas},\ }\href@noop {} {\bibfield  {journal} {\bibinfo  {journal}
  {Science}\ }\textbf {\bibinfo {volume} {370}},\ \bibinfo {pages} {1222}
  (\bibinfo {year} {2020})}\BibitemShut {NoStop}%
\bibitem [{\citenamefont {Wang}\ \emph {et~al.}(2022)\citenamefont {Wang},
  \citenamefont {Li}, \citenamefont {Arakelyan},\ and\ \citenamefont
  {Thomas}}]{XinHydroRelax}%
  \BibitemOpen
  \bibfield  {author} {\bibinfo {author} {\bibfnamefont {X.}~\bibnamefont
  {Wang}}, \bibinfo {author} {\bibfnamefont {X.}~\bibnamefont {Li}}, \bibinfo
  {author} {\bibfnamefont {I.}~\bibnamefont {Arakelyan}},\ and\ \bibinfo
  {author} {\bibfnamefont {J.~E.}\ \bibnamefont {Thomas}},\ }\bibfield  {title}
  {\bibinfo {title} {Hydrodynamic relaxation in a strongly interacting {Fermi}
  gas},\ }\href@noop {} {\bibfield  {journal} {\bibinfo  {journal} {Phys. Rev.
  Lett.}\ }\textbf {\bibinfo {volume} {128}},\ \bibinfo {pages} {090402}
  (\bibinfo {year} {2022})}\BibitemShut {NoStop}%
\bibitem [{\citenamefont {Li}\ \emph {et~al.}(2022)\citenamefont {Li},
  \citenamefont {Luo}, \citenamefont {Wang}, \citenamefont {Xie}, \citenamefont
  {Liu}, \citenamefont {Hu}, \citenamefont {Chen}, \citenamefont {Yao},\ and\
  \citenamefont {Pan}}]{SecondSoundLi}%
  \BibitemOpen
  \bibfield  {author} {\bibinfo {author} {\bibfnamefont {X.}~\bibnamefont
  {Li}}, \bibinfo {author} {\bibfnamefont {X.}~\bibnamefont {Luo}}, \bibinfo
  {author} {\bibfnamefont {S.}~\bibnamefont {Wang}}, \bibinfo {author}
  {\bibfnamefont {K.}~\bibnamefont {Xie}}, \bibinfo {author} {\bibfnamefont
  {X.-P.}\ \bibnamefont {Liu}}, \bibinfo {author} {\bibfnamefont
  {H.}~\bibnamefont {Hu}}, \bibinfo {author} {\bibfnamefont {Y.-A.}\
  \bibnamefont {Chen}}, \bibinfo {author} {\bibfnamefont {X.-C.}\ \bibnamefont
  {Yao}},\ and\ \bibinfo {author} {\bibfnamefont {J.-W.}\ \bibnamefont {Pan}},\
  }\bibfield  {title} {\bibinfo {title} {Second sound attenuation near quantum
  criticality},\ }\href@noop {} {\bibfield  {journal} {\bibinfo  {journal}
  {Science}\ }\textbf {\bibinfo {volume} {375}},\ \bibinfo {pages} {528}
  (\bibinfo {year} {2022})}\BibitemShut {NoStop}%
\bibitem [{\citenamefont {Hu}\ \emph {et~al.}(2018)\citenamefont {Hu},
  \citenamefont {Zou},\ and\ \citenamefont {Liu}}]{HuTwoFluidPRA}%
  \BibitemOpen
  \bibfield  {author} {\bibinfo {author} {\bibfnamefont {H.}~\bibnamefont
  {Hu}}, \bibinfo {author} {\bibfnamefont {P.}~\bibnamefont {Zou}},\ and\
  \bibinfo {author} {\bibfnamefont {X.-J.}\ \bibnamefont {Liu}},\ }\bibfield
  {title} {\bibinfo {title} {Low-momentum dynamic structure factor of a
  strongly interacting {Fermi} gas at finite temperature: A two-fluid
  hydrodynamic description},\ }\href@noop {} {\bibfield  {journal} {\bibinfo
  {journal} {Phys. Rev. A}\ }\textbf {\bibinfo {volume} {97}},\ \bibinfo
  {pages} {023615} (\bibinfo {year} {2018})}\BibitemShut {NoStop}%
\bibitem [{\citenamefont {Yan}\ \emph {et~al.}(2024)\citenamefont {Yan},
  \citenamefont {Patel}, \citenamefont {Mukherjee}, \citenamefont {Vale},
  \citenamefont {Fletcher},\ and\ \citenamefont {Zwierlein}}]{MZTempWave}%
  \BibitemOpen
  \bibfield  {author} {\bibinfo {author} {\bibfnamefont {Z.}~\bibnamefont
  {Yan}}, \bibinfo {author} {\bibfnamefont {P.~B.}\ \bibnamefont {Patel}},
  \bibinfo {author} {\bibfnamefont {B.}~\bibnamefont {Mukherjee}}, \bibinfo
  {author} {\bibfnamefont {C.~J.}\ \bibnamefont {Vale}}, \bibinfo {author}
  {\bibfnamefont {R.~J.}\ \bibnamefont {Fletcher}},\ and\ \bibinfo {author}
  {\bibfnamefont {M.~W.}\ \bibnamefont {Zwierlein}},\ }\bibfield  {title}
  {\bibinfo {title} {Thermography of the superfluid transition in a strongly
  interacting {Fermi} gas},\ }\href@noop {} {\bibfield  {journal} {\bibinfo
  {journal} {Science}\ }\textbf {\bibinfo {volume} {383}},\ \bibinfo {pages}
  {629} (\bibinfo {year} {2024})}\BibitemShut {NoStop}%
\bibitem [{\citenamefont {Son}(2007)}]{SonBulkViscosity}%
  \BibitemOpen
  \bibfield  {author} {\bibinfo {author} {\bibfnamefont {D.~T.}\ \bibnamefont
  {Son}},\ }\bibfield  {title} {\bibinfo {title} {Vanishing bulk viscosities
  and conformal invariance of the unitary {Fermi} gas},\ }\href@noop {}
  {\bibfield  {journal} {\bibinfo  {journal} {Phys. Rev. Lett.}\ }\textbf
  {\bibinfo {volume} {98}},\ \bibinfo {pages} {020604} (\bibinfo {year}
  {2007})}\BibitemShut {NoStop}%
\bibitem [{\citenamefont {Hou}\ \emph {et~al.}(2013)\citenamefont {Hou},
  \citenamefont {Pitaevskii},\ and\ \citenamefont {Stringari}}]{StringariBulk}%
  \BibitemOpen
  \bibfield  {author} {\bibinfo {author} {\bibfnamefont {Y.-H.}\ \bibnamefont
  {Hou}}, \bibinfo {author} {\bibfnamefont {L.~P.}\ \bibnamefont
  {Pitaevskii}},\ and\ \bibinfo {author} {\bibfnamefont {S.}~\bibnamefont
  {Stringari}},\ }\bibfield  {title} {\bibinfo {title} {Scaling solutions of
  the two-fluid hydrodynamic equations in a harmonically trapped gas at
  unitarity},\ }\href@noop {} {\bibfield  {journal} {\bibinfo  {journal} {Phys.
  Rev. A}\ }\textbf {\bibinfo {volume} {87}},\ \bibinfo {pages} {033620}
  (\bibinfo {year} {2013})}\BibitemShut {NoStop}%
\bibitem [{\citenamefont {Elliott}\ \emph {et~al.}(2014)\citenamefont
  {Elliott}, \citenamefont {Joseph},\ and\ \citenamefont
  {Thomas}}]{ElliottScaleInv}%
  \BibitemOpen
  \bibfield  {author} {\bibinfo {author} {\bibfnamefont {E.}~\bibnamefont
  {Elliott}}, \bibinfo {author} {\bibfnamefont {J.~A.}\ \bibnamefont
  {Joseph}},\ and\ \bibinfo {author} {\bibfnamefont {J.~E.}\ \bibnamefont
  {Thomas}},\ }\bibfield  {title} {\bibinfo {title} {Observation of conformal
  symmetry breaking and scale invariance in expanding {Fermi} gases},\
  }\href@noop {} {\bibfield  {journal} {\bibinfo  {journal} {Phys. Rev. Lett.}\
  }\textbf {\bibinfo {volume} {112}},\ \bibinfo {pages} {040405} (\bibinfo
  {year} {2014})}\BibitemShut {NoStop}%
\bibitem [{\citenamefont {Ku}\ \emph {et~al.}(2012)\citenamefont {Ku},
  \citenamefont {Sommer}, \citenamefont {Cheuk},\ and\ \citenamefont
  {Zwierlein}}]{KuThermo}%
  \BibitemOpen
  \bibfield  {author} {\bibinfo {author} {\bibfnamefont {M.}~\bibnamefont
  {Ku}}, \bibinfo {author} {\bibfnamefont {A.~T.}\ \bibnamefont {Sommer}},
  \bibinfo {author} {\bibfnamefont {L.~W.}\ \bibnamefont {Cheuk}},\ and\
  \bibinfo {author} {\bibfnamefont {M.~W.}\ \bibnamefont {Zwierlein}},\
  }\bibfield  {title} {\bibinfo {title} {Revealing the superfluid lambda
  transition in the universal thermodynamics of a unitary {Fermi} gas},\
  }\href@noop {} {\bibfield  {journal} {\bibinfo  {journal} {Science}\ }\textbf
  {\bibinfo {volume} {335}},\ \bibinfo {pages} {563} (\bibinfo {year}
  {2012})}\BibitemShut {NoStop}%
\bibitem [{Sup()}]{SupportOnline}%
  \BibitemOpen
  \href@noop {} {}\bibinfo {note} {See the Supplemental Material for discussion
  of the linearized hydrodynamic equations, the kinetic theory relaxation
  model, and the determination of the static transport properties.}\BibitemShut
  {Stop}%
\bibitem [{Dru()}]{Drude}%
  \BibitemOpen
  \href@noop {} {}\bibinfo {note} {In a simple Drude model with a transport
  time $\tau$, the electrical conductivity is
  $\sigma(\omega)=\sigma(0)/(1-i\omega\tau)$.}\BibitemShut {Stop}%
\bibitem [{PiF()}]{PiFactor}%
  \BibitemOpen
  \href@noop {} {}\bibinfo {note} {Note that the Fermi time defined in
  Ref.~\cite{ZwergerViscosity}, Fig.~6 is $\hbar/\epsilon_F$, which is a factor
  of $\pi$ smaller than $\lambda_F/v_F$.}\BibitemShut {Stop}%
\bibitem [{Tra()}]{TransportError}%
  \BibitemOpen
  \href@noop {} {}\bibinfo {note} {The vertical error bars in Figs.~3~and~4
  denote $\pm\sqrt{2\epsilon_{ii}}$, where $\epsilon_{ij}$ is the error matrix
  obtained from $\chi^2(\tau_\eta,\tau_\kappa)$ with $A$ and $c_T$
  fixed.}\BibitemShut {Stop}%
\bibitem [{\citenamefont {Landau}\ and\ \citenamefont
  {Lifshitz}(1959)}]{LandauFluids}%
  \BibitemOpen
  \bibfield  {author} {\bibinfo {author} {\bibfnamefont {L.~D.}\ \bibnamefont
  {Landau}}\ and\ \bibinfo {author} {\bibfnamefont {E.~M.}\ \bibnamefont
  {Lifshitz}},\ }\href@noop {} {\emph {\bibinfo {title} {Fluid Dynamics, Course
  of Theoretical Physics Vol. VI}}}\ (\bibinfo  {publisher} {Pergamon Press,
  Oxford},\ \bibinfo {year} {1959})\BibitemShut {NoStop}%
\end{thebibliography}
\end{document}